\documentclass[aps,pra,twocolumn,showpacs,superscriptaddress,longbibliography]{revtex4-1}
\usepackage{amsmath,braket,amssymb,natbib,graphicx,float,amsthm,xcolor,outlines,mathrsfs,multirow,hhline,bm,mathtools}

\usepackage{dcolumn}% Align table columns on decimal point
\usepackage[english]{babel}
\usepackage[utf8]{inputenc}
\usepackage[T1]{fontenc}
\usepackage{listings}
\usepackage[colorinlistoftodos]{todonotes} 
\usepackage[colorlinks=true, allcolors=blue]{hyperref}
\usepackage[caption=false]{subfig}
\begin{document}
\title{Arbitrary high-fidelity binomial codes from multiphoton spin-boson interactions}
% Authors
\author{Pradip Laha}
\email{plaha@uni-mainz.de}
\affiliation{Institute of Physics, Johannes Gutenberg-Universit\"at Mainz, Staudingerweg 7, 55128 Mainz, Germany}

\author{Peter van Loock}
\email{loock@uni-mainz.de}
\affiliation{Institute of Physics, Johannes Gutenberg-Universit\"at Mainz, Staudingerweg 7, 55128 Mainz, Germany}

\begin{abstract}
Encoding a qubit in the continuous degrees of freedom of a quantum system, such as bosonic modes, is a powerful alternative to modern quantum error correction (QEC). Among the most prominent bosonic QEC codes, binomial codes provide protection against loss and dephasing errors by encoding logical states in finite superpositions of Fock states with binomially weighted coefficients. While much attention has been given to their error-correcting capabilities and integration into fault-tolerant architectures, efficient methods for generating arbitrary binomial codewords remain scarce. In this work, we propose a scheme for generating these codewords by exploiting nonlinear multiphoton interactions between a continuous-variable bosonic mode (oscillator) and a two-level system (spin/qubit). Our proposed scheme assumes the ability to prepare the oscillator in an arbitrary Fock state and the qubit in an arbitrary superposition of its basis states and access to arbitrarily high multiphoton interactions. To enhance the experimental feasibility of our scheme, we further demonstrate how to reduce the required order parameter of multiphoton interactions by a factor of two for a special class of code states.
\end{abstract}

\maketitle
\section{Introduction}
Quantum error correction (QEC) plays a foundational role in enabling scalable quantum information processing, providing a systematic framework to safeguard fragile quantum states from decoherence and other operational noise~\cite{nielsen_chuang_2010,Devitt_2013,Terhal_RMP_2015,Roffe_2019,Cho_Science_2020}. While conventional QEC schemes typically encode logical information into discrete-variable qubit registers, continuous-variable (CV) approaches leverage the infinite-dimensional Hilbert space structure of bosonic modes (or qumodes, such as quantized electromagnetic fields) to encode and protect quantum information~\cite{Braunstein_RMP_2005,Weedbrook_RMP_2012,Serafini_2017}. Owing to their intrinsic redundancy and potential for resource-efficient error mitigation~\cite{Braunstein_PRL_1998}, CV-based error correction has attracted growing interest, making bosonic codes particularly promising for near-term quantum hardware where qubit resources are limited and noise is prevalent~\cite{Albert_PRA_2018,Fluhmann_nature_2019,Hu_NP_2019,Terhal_QST_2020,Grimsmo_PRX_2020,Cai_FR_2021,Joshi_QST_2021,Ma_SB_2021,Brennan_nat_phys_2022,Wang_PRXQ_2022,Sivak_nature_2023,Brady_PQE_2024,xu_2024}.
Prominent examples of encoding a qubit in bosonic modes include Gottesman-Kitaev-Preskill (GKP) codes~\cite{GKP_PRA_2001,Noh_IEEE_2019,Baragiola_PRL_2019,Noh_PRA_2020,Noh_PRXQ_2022,Conrad_2024}, which protect against small displacement errors, and cat codes~\cite{Cochrane_PRA_1999,Leghtas_PRL_2013,Mirrahimi_NJP_2014,Bergmann_PRA_2016,Li_PRL_2017,Chamberland_PRXQ_2022}, which leverage coherent state superpositions to mitigate photon loss. These code states have demonstrated remarkable potential in experimental platforms such as superconducting circuits~\cite{Ofek_nature_2016,Heeres_nat_com_2017,Hu_NP_2019,Campagne_nat_2020}, optical systems~\cite{Dahan_PRX_2023,Konno_science_2024,Chen_PRA_2024},  and trapped ions~\cite{Blatt_nature_2008,Fluhmann_nature_2019,Brennan_nat_phys_2022}. 

Alongside these, binomial encoding~\cite{Stoler_1985} has emerged as a compelling alternative, offering protection against both photon loss and dephasing errors while requiring only modest hardware overhead~\cite{Albert_PRX_2016}. These codes encode quantum information into {\em finite} superpositions of Fock states with binomially weighted amplitudes, carefully designed so that common error processes map codewords onto orthogonal, correctable states. Their discrete structure, analytical tractability, and compatibility with bosonic modes make them highly attractive for near-term quantum technologies~\cite{Hu_NP_2019,Chen_PRR_2021,Wollack_nature_2022,Kang_OL_2022,Ni_nature_2023,xu_2024,Juliette_2024,Li_QST_2025,chang_2025}. Binomial codes have been explored across a range of physical architectures, including superconducting circuits~\cite {Hu_NP_2019,Wollack_nature_2022,Ni_nature_2023}, light-matter systems~\cite{Chen_PRR_2021}, and trapped ions~\cite{Fluhmann_nature_2019}. Recent studies have evaluated the fault tolerance and scalability of binomial codes under realistic noise~\cite{udupa_2025}. Reference~\cite{Juliette_2024} examined concatenated architectures and identified photon-loss thresholds for different measurement strategies, while Ref.~\cite{tanaka_2024} proposes a scheme for logical qubit rotations within the binomial code space without the need for an ancillary qubit. Further, a quantum repeater architecture based on binomial codes has been proposed, employing a microwave cavity and a superconducting transmon to suppress photon loss while enabling high-fidelity state preparation and entanglement swapping~\cite{Siddhu_2025}. In parallel, high-fidelity preparation of binomial code states has been experimentally demonstrated in a circuit-quantum electrodynamics (QED) architecture by employing transmon-mediated sideband interactions, enabling efficient multimode control and state encoding~\cite{huang_2025}.

%\cite{chang2025}
\begin{figure*}
    \includegraphics[width=1.0\textwidth]{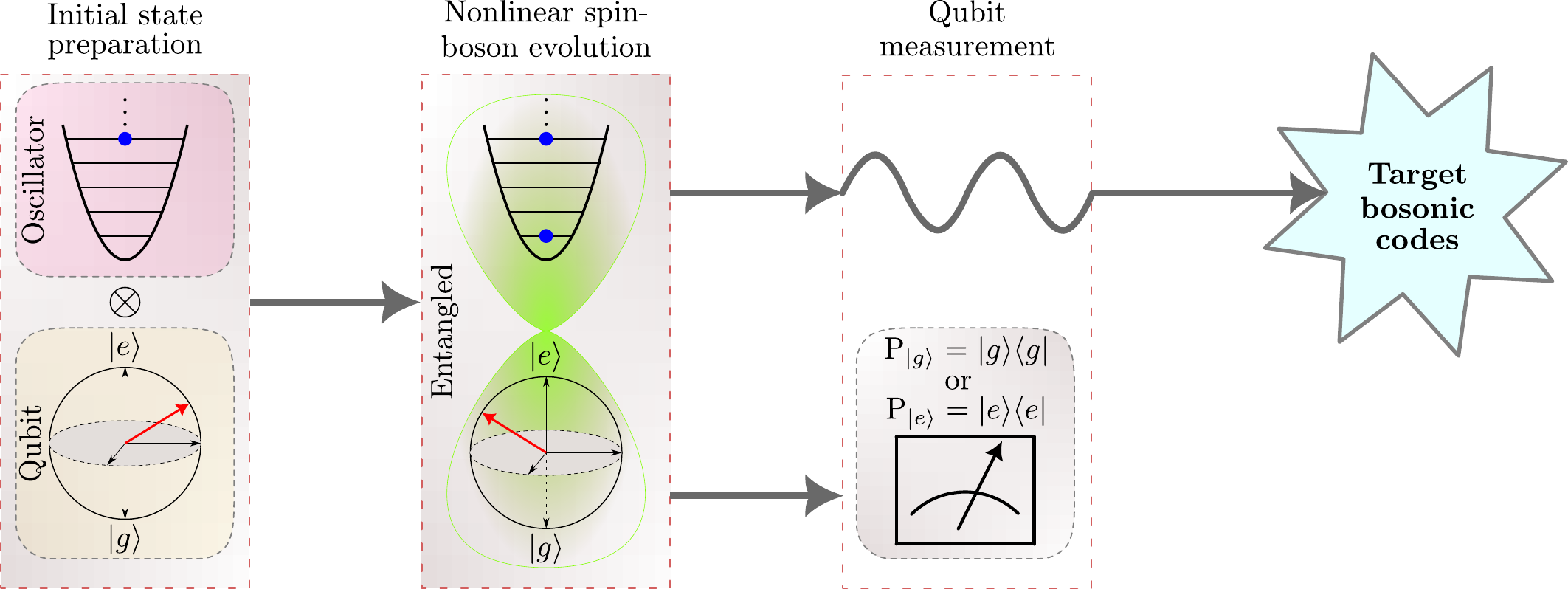}
    \vspace{-2ex}
    \caption{Schematic of the proposed protocol for synthesizing binomial code states composed of two Fock state superpositions. The protocol begins with state preparation, where the oscillator is initialized in a Fock state $\ket{n_1}$ (blue solid circle) and the qubit in a superposition state $\cos\theta\ket{g}+\sin\theta\ket{e}$ (red arrow). The joint system then undergoes an entangling evolution governed by a nonlinear multiphoton Jaynes–Cummings (MPJC) interaction of order $m$, mediated by a tunable coupling strength $g$. A projective measurement on the qubit at an optimized interaction time $\tau$ collapses the oscillator into a nontrivial superposition of Fock states.
    %A final local unitary operation on the oscillator corrects the relative phase and completes the synthesis of the desired binomial code state.
    The parameters $n_1$, $m$, $\theta$, and $\tau$ uniquely determine the final encoded binomial state. The protocol can be recursively applied to construct more complex binomial states involving additional Fock components (see main text).}
    \label{fig:schematic}
\end{figure*}

Despite significant progress, the high-fidelity generation of {\em arbitrary} binomial codewords remains a critical challenge. Most recent efforts to date have primarily focused on the simplest instance—protection against a single photon loss—where the two logical states of the qubit
%$\ket{\bar{0}}$ and $\ket{\bar{1}}$
are given by $\ket{\bar{0}} = (\ket{0} + \ket{4})/\sqrt{2}$ and $\ket{\bar{1}} = \ket{2}$, respectively. Here, $\ket{n}$ is an $n-$photon Fock state ($n=0, 2, 4$). However, the general binomial code framework introduced in Ref.~\cite{Albert_PRX_2016} enables protection against multiple photon loss, gain, and dephasing errors, albeit with increasingly complex logical encodings. Developing a universal, hardware-independent protocol for the high-fidelity preparation of arbitrary binomial codewords would mark a major step toward scalable, modular quantum architectures based on bosonic encodings.

In this work, we address this gap by introducing a protocol based on the multiphoton Jaynes–Cummings (MPJC) interaction, a highly nonlinear coupling between a two-level system (spin/qubit) and a bosonic mode. Initially developed in the context of nonlinear optics and cavity QED~\cite{SUKUMAR_pla_1981,surendra_pra_1982,SHUMOVSKY_pla_1987,Kien_PRA_1988,Vogel_PRA_1989}, MPJC-type Hamiltonians have seen renewed interest~\cite{Villas_PRL_2019,Puebla_symm_2019,Mavrogordatos_PRA_2021,laha_PRR_2024,laha_PRA_2025} due to their increasing experimental accessibility in quantum platforms such as superconducting circuits~\cite{Strauch2010,Felicetti_PRA_2018} and trapped ions~\cite{Leibfried_RMP_2003,Haffner_phys_rep_2008}. 

Our approach exploits the nonlinearities inherent in the multiphoton generalization of the standard Jaynes–Cummings (JC) interaction to coherently drive the oscillator into arbitrary binomial codewords during its time evolution with high fidelity.
Note that the standard JC interaction as directly available in cavity QED, including its dispersive regimes enabling spin-controlled phase rotations of an oscillator mode, has been shown~\cite{Li_AdvQTech_2023} to be applicable for both cat code state engineering and cat code photon loss error syndrome identification. For the former, standard cavity QED interactions can be employed in an iterative fashion, starting from a {\em primitive state}~\cite{Grimsmo_PRX_2020}. It is tempting to consider a similar iterative approach to the generation of binomial code states. However, while the primitive state for the cat code is a simple coherent state, the primitive states associated with the binomial code are, from an experimental point of view, not really easier to realize than the target codewords themselves (see Appendix~\ref{sce:RSBC} for a brief discussion on this).

%\cite{Hastrup_PRL_2022}

%The standard JC interaction, readily available in cavity QED systems—including its dispersive regimes that enable spin-controlled phase rotations—has been successfully applied to both the generation of cat code states and the detection of photon loss error syndromes~\cite{Li_AdvQTech_2023}. In particular, cat code states can be prepared iteratively using conventional cavity QED interactions, beginning with a simple coherent state as a primitive state~\cite{Grimsmo_PRX_2020}. This raises the question of whether a similar iterative scheme could be employed for generating binomial codewords. However, while the coherent-state primitive of the cat code is straightforward to prepare experimentally, the primitive states associated with binomial codes are more complex superpositions of Fock states and, in practice, are not easier to realize than the codewords themselves. As discussed in Appendix~\ref{sce:RSBC}, this limits the practical advantage of rotation-based constructions for binomial code state preparation.

To achieve our objective, we assume initialization of the oscillator in a known Fock state and the qubit in an arbitrary superposition of its two basis states and access to tunable multiphoton interactions of arbitrary order. Under these conditions, and with appropriate local unitaries, we demonstrate high-fidelity synthesis of any binomial codeword (see Fig.~\ref{fig:schematic}). While our primary focus is on the probabilistic protocol—based on postselected qubit measurements—owing to its enhanced flexibility in parameter selection and higher achievable fidelities, we also demonstrate that similar target states can be synthesized deterministically by tracing out the qubit, albeit with a modest reduction in fidelity due to the mixedness introduced by this operation. Both analytical and numerical results confirm the scalability of our protocol to increasingly complex logical states.

It is instructive to contrast our protocol with schemes for GKP state generation in cavity and circuit QED, which typically rely on sequences of distinct interactions to engineer grid states in oscillator quadratures~\cite{Hastrup_PRL_2022}. In contrast, our approach achieves deterministic preparation of structured bosonic codewords by tuning only the interaction time of a single MPJC Hamiltonian. While the resulting states form a comb in photon-number space rather than quadrature space, both protocols exploit controlled nonlinear light–matter interactions to generate highly nonclassical oscillator states.

While arbitrary oscillator Fock state preparation is now feasible across many quantum platforms~\cite{Meekhof_PRL_1996,Hofheinz_nature_2008,Chu_nature_2018,Wolf_NC_2019,Uwe_nat_phys_2022,Li_PRL_2024,Deng_nat_phys_2024,Rahman_PRL_2025}, it is crucial to note that neither the standard nor multiphoton JC interactions can generate subsystem coherence from an initially incoherent joint state~\cite{Vogel_PRA_1989,laha_AdvQT_2023,laha_PRR_2024}. In our protocol, coherence in the spin state is therefore essential, as it seeds the coherence required in the initially incoherent oscillator state. Optimizing this initial spin coherence is consequently key to synthesizing the target bosonic state in a single step with high fidelity. A practical challenge in multiphoton protocols is the scaling of the required multiphoton order parameter with increasing code distance. To improve feasibility, we propose a two-step implementation that reduces the necessary photon-number requirement by a factor of two. This one-time improvement is achieved through judicious choices of initial conditions and by leveraging symmetries in the MPJC dynamics. Finally, to analyze the robustness of our scheme in realistic settings, we also briefly discuss the role of inevitable environmental-induced effects, including oscillator damping and qubit dephasing. 

The remainder of this paper is organized as follows. Section~\ref{sec_binom_states} provides a short overview of the structure and properties of binomial codes. In Sec.~\ref{sec_MPJC}, we briefly introduce the MPJC Hamiltonian and describe the theoretical framework of our state-generation protocol. Sections~\ref{sec_two_fock} and~\ref{sec_three_fock} present explicit constructions of binomial codewords involving two and three Fock components, respectively, using a probabilistic scheme. In Sec.~\ref{sec_reduced_m},  we examine how the required multiphoton parameter can be reduced, improving experimental feasibility. Section~\ref{sec_env} analyzes the impact of decoherence due to system–environment coupling on the fidelity of the generated states. The deterministic state generation protocol is discussed in Sec.~\ref{sec_deterministic}. We conclude in Sec.~\ref{sec_conc} with a brief summary and outlook. Supplementary technical details and extended results are provided in the appendixes.

\section{Binomial code states}
\label{sec_binom_states}
As noted earlier, the generic binomial code encodes the two logical qubit states into \emph{finite} superpositions of Fock states with binomially distributed amplitudes. It was first introduced in Ref.~\cite{Albert_PRX_2016} as a two-parameter ($N, K$) code space to correct against up to $L$ photon losses, up to $G$ photon gain errors, and $D$ dephasing events. Here, $N=L+G+1$ decides the spacing of the superpositions of Fock states, and $K=\text{max}\{L, G, 2D\}+1$ is responsible for the number of Fock states that are present in the superposition.
 The generic class of the two logical states is defined as 
%\begin{align}
%    \ket{\bar{\mu}}_{N,S} = \frac{1}{\sqrt{2^N}}\sum_{k=0}^{\left\lceil\frac{{N+1}}{2}\right\rceil-\mu}\sqrt{\binom{N+1}{2k+\mu}} \ket{(S+1)(2k+\mu)},
%    \label{eqn_bonom_code_states}
%\end{align}
\begin{subequations}
    \label{eqn_bonom_code_states}
\begin{align}
    \label{eqn_bonom_code_state_0}
    \ket{\bar{0}}_{N,K} &= \frac{1}{\sqrt{2^{K-1}}}\sum_{k=0}^{\left\lfloor\frac{{K}}{2}\right\rfloor}\sqrt{\binom{K}{2k}} \ket{2kN)},\\
    \ket{\bar{1}}_{N,K} &= \frac{1}{\sqrt{2^{K-1}}}\sum_{k=0}^{\left\lfloor\frac{{K-1}}{2}\right\rfloor}\sqrt{\binom{K}{(2k+1)}} \ket{(2k+1)N)},
    \label{eqn_bonom_code_state_1}
\end{align}
\end{subequations}
where, $\lfloor \cdot \rfloor$ denotes the conventional floor function~\cite{Juliette_2024}. For simplicity, we neglect photon gain errors in this work (i.e., $G = 0$), such that $N = L+1$ and $K= \max\{L, 2D\}+1$. For a fixed value of $L$, the number of correctable dephasing errors is determined by $K$. In the asymptotic limit $K \to \infty$, the binomial codes asymptotically approach the $2(L+1)$-legged cat codes~\cite{Leghtas_PRL_2013,Mirrahimi_NJP_2014,Bergmann_PRA_2016,Li_PRL_2017}, which are known to protect against up to $L$ photon loss errors, as discussed in Ref.~\cite{Albert_PRX_2016}. Both binomial and cat codes belong to the broader class of rotation-symmetric bosonic codes (RSBCs)~\cite{Grimsmo_PRX_2020, Endo_PRA_2025, udupa_2025}, characterized by their invariance under discrete rotations by a fixed angle in phase space. In this context,  the parameter $N$ quantifies the degree of discrete rotation symmetry for the code.
%We now illustrate several explicit examples of binomially encoded logical qubit states.

The simplest and most well-known instance—mentioned in the previous section—corresponds to the case $L=1$, which corrects a single photon loss. The associated logical states are $ \ket{\bar{0}}_{2,2} =  \tfrac{1}{\sqrt{2}} \left(\ket{0} + \ket{4}\right)$, and $\ket{\bar{1}}_{2,2} =  \ket{2}$, where $K=2$. By fixing $L=1$ and increasing $K$, one obtains codewords that maintain protection against single photon loss while improving resilience to dephasing. On the other hand, the lowest-order binomial codewords capable of correcting up to two photon loss events ($L=2$) are $ \ket{\bar{0}}_{3,3} =  \tfrac{1}{2} \left(\ket{0} + \sqrt{3}\ket{6}\right)$ and $ \ket{\bar{1}}_{3,3} =  \tfrac{1}{2} \left(\sqrt{3}\ket{3} + \ket{9}\right)$.
%\begin{subequations}
%\begin{align}
%  \ket{\bar{0}}_{3,3} &=  \tfrac{1}{2} \left(\ket{0} + \sqrt{3}\ket{6}\right), \\
%  \ket{\bar{1}}_{3,3} &=  \tfrac{1}{2} \left(\sqrt{3}\ket{3} + \ket{9}\right).
% \end{align}
%\label{eqn_bcs_N2S2}
%\end{subequations}
Higher-order encoded states corresponding to larger $N$ and $K$ values can be systematically constructed from Eq.~\eqref{eqn_bonom_code_states}. The objective of this work is to produce arbitrary superpositions of these Fock states with high fidelity.

%%%%%%%%%%%%%%%%%%%%%%%%%%%%%%%%%%%%%%%%%%%%%%%%%%%%%%%%%%%%%%%%%%%%%%%%%%%%%%%%%%%%

\section{Multiphoton spin-boson Hamiltonian}
\label{sec_MPJC}
The JC model~\cite{jaynes_cummings_1963,Shore_JCM_1993} captures the fundamental physics of light–matter interactions and forms the basis for understanding cavity and circuit QED systems. It plays a central role in the theoretical and experimental development of quantum optics and quantum information platforms~\cite{Larson_JCM_2021}. The standard JC model describes the coherent coupling between a two-level system (spin) and a single quantized harmonic oscillator mode (boson). The two states of the qubit, $\ket{g}$ and $\ket{e}$, are assumed to be separated by a transition frequency $\omega_0$, while the oscillator is characterized by respective bosonic creation and annihilation operators $a^\dagger$ and $a$, with mode frequency $\omega$. The detuning is typically denoted as $\Delta = \omega_0-\omega$. The free energy Hamiltonian is expressed as $H_0 = \frac{\omega_0}{2}\sigma_z + \omega a^\dagger a$ while the interaction Hamiltonian is given by $H_{\text{int}} = \textsl{g}(a\sigma_+ + a^{\dagger}\sigma_-)$ where $ \textsl{g}$ denotes the strength of the spin-boson interaction. Here, $\sigma_z = \ket{e}\bra{e}-\ket{g}\bra{g}$, $\sigma_+=\ket{e}\bra{g}$, $\sigma_-=\ket{g}\bra{e}$ are the standard spin operators. 
The total Hamiltonian is therefore $H=H_0+H_{\text{int}}$. It is often useful to split $H$ into two mutually commuting terms as $H=H_I+H_{II}$, where $H_I = \omega\left(a^\dagger a + \frac{\sigma_z}{2}\right)$ and $H_{II} = \frac{\Delta}{2}\sigma_z+H_{\text{int}}$. In the limit $\Delta = 0$ (on resonance), $H_{II} \equiv H_{\text{int}}$.

Over the years, the standard JC model has been extended in various directions to capture richer quantum phenomena (see Ref.~\cite{Larson_JCM_2021} for a review). Among these, the multiphoton generalization is particularly notable~\cite{SUKUMAR_pla_1981,surendra_pra_1982,SHUMOVSKY_pla_1987,Kien_PRA_1988,Villas_PRL_2019}. The  MPJC model describes a coherent exchange of $m$ bosonic excitations with a two-level system. Its interaction Hamiltonian is given by 
\begin{align}
    H_{\text{int}} =  i \textsl{g}(a^m\sigma_+ - a^{\dagger\, m}\sigma_-),
    \label{eqn_H_int}
\end{align} 
where $m$ is the multiphoton order parameter and $\textsl{g}$ is the effective coupling strength. 
Note that the interaction Hamiltonian $H_{\text{int}}$ in Eq.~\eqref{eqn_H_int} is equivalent to the conventional form $H'_{\text{int}} = g(a^{m}\sigma_{+} + a^{\dagger m}\sigma_{-})$, differing only by a phase convention for the two-level system operators and, correspondingly, for the coupling constant. Specifically, applying the transformation $\sigma_{-} \rightarrow e^{i\phi}\sigma_{-}$ (and $\sigma_{+} \rightarrow e^{-i\phi}\sigma_{+}$) in $H'_{\text{int}}$ and choosing $\phi = \pi/2$ recovers $H_{\text{int}}$.
The case $m = 1$ corresponds to the standard JC model. In what follows, we will find out that $m$ plays a crucial role in engineering the desired quantum superposition states.  Throughout this work, we restrict ourselves to the resonant regime $\Delta = \omega_0-m\omega = 0$, where the MPJC dynamics admits a transparent analytical solution and directly enables the state-engineering protocols discussed below. Extensions of the protocol to finite detuning $\Delta \neq 0$ are in principle possible but require a separate analysis and are beyond the scope of the present work.

We mention in passing a recent circuit QED demonstration of single-mode control over cavity modes by preparing both Fock states $\ket{n}$ and vacuum–Fock superpositions $(\ket{0}+\ket{n})/\sqrt{2}$ by climbing the JC ladder with transmon rotations and sideband pulses. Fock states are prepared using a sequence of broadband pulses followed by sideband $\pi$ pulses. Consequently, a shelving method is employed to achieve superposition states~\cite{huang_2025}.

\subsection{Time-evolved state of the MPJC Hamiltonian}
Before proceeding, we summarize the key assumptions of our state preparation protocol, as introduced earlier: (i) the oscillator is initialized in a Fock state $\ket{n_1}$, while the qubit starts in an arbitrary superposition $\ket{\psi_q(\theta)}=\cos\theta\ket{g} + \sin\theta\ket{e}$); (ii) tunable access to MPJC interactions ($H_{\text{int}}$ in Eq.~\eqref{eqn_H_int}) with arbitrary $m$ is available; and (iii) projective qubit measurements can be performed at chosen times during the evolution dictated by the target state. 
Under these assumptions, the binomial target state preparation protocol can be succinctly summarized as\begin{equation}
\ket{\psi_o}_{\mathrm{target}}\;\propto\;
\bra{g/e}\, e^{-i H_{\mathrm{int}} \tau}
\bigl(\ket{\psi_q(\theta)} \otimes \ket{\psi_o(0)}\bigr).
\label{eq:protocol_summary}
\end{equation}
Notably, we will later show that employing a two-step generation protocol—rather than a single-step approach—can reduce the required value of $m$ by a factor of two, significantly easing experimental demands. 

According to the first assumption, the initial joint state of the system can be written as
\begin{align}
    \ket{\Psi(0)} &=\cos\theta\ket{g, n_1}+\sin\theta\ket{e, n_1},
    \label{eqn_psi0}
\end{align}
where the notation is self-evident.
It is important to emphasize that neither the standard JC Hamiltonian nor its multiphoton generalization (MPJC) can generate coherence within either subsystem if the joint system initially lacks coherence~\cite{laha_AdvQT_2023,laha_PRR_2024}. In the case of the spin, this is because $H_{\text{int}}$ merely rotates the Bloch vector trivially along the $z$-axis~\cite{laha_AdvQT_2023}. As we will demonstrate, initial coherence in the qubit (parametrized by the angle $\theta$) is essential for generating the desired superpositions in the oscillator.

We now turn to the time-evolved state vector $\ket{\psi(t)}$, governed by the interaction Hamiltonian $H_\text{int}$ in Eq.~\eqref{eqn_H_int}, starting from a pure initial product state $\ket{\psi(0)}$ in Eq.~\eqref{eqn_psi0}. It is straightforward to show that the evolved state has the form
\begin{align}
    \ket{\Psi(\tau)} = x_1(\tau)\ket{g, n_1} &+ x_2(\tau) \ket{e, n_1} + x_3(\tau)\ket{g, n_1+m} \nonumber \\
    &\quad\quad+ x_4(\tau)\ket{e, n_1-m},  
    \label{eqn_psit_2fock}
\end{align}
where the time-dependent coefficients are found to be
\begin{subequations}
\begin{align}
    x_1(\tau) &= \cos\theta \cos\left( \sqrt{n_1!/(n_1-m)!}\, \tau\right), \\
    x_2(\tau) &= \sin\theta \cos \left(\sqrt{(n_1+m)!/n_1!}\, \tau\right), \\
    x_3(\tau) &= -\sin\theta \sin \left(\sqrt{(n_1+m)!/n_1!}\, \tau\right), \\
    x_4(\tau) &= \cos\theta \sin\left( \sqrt{n_1!/(n_1-m)!} \, \tau\right),
\end{align}    
\label{eqn_xt_2fock}
\end{subequations}
for $n_1\geqslant m$. When $n_1<m$, $x_1(\tau) = \cos\theta, \quad  x_4(\tau) = 0$.
%\begin{subequations}
%\begin{align}
%    x_1(\tau) = \cos\theta, \quad  x_4(\tau) = 0.
%\end{align}    
%\end{subequations}
 Here,  $\tau= \textsl{g}\,t$ denotes the dimensionless scaled interaction time (see Appendix~\ref{sec_app_state_vector} for details).

\begin{figure*}
    \centering
    \includegraphics[width=1.0\linewidth]{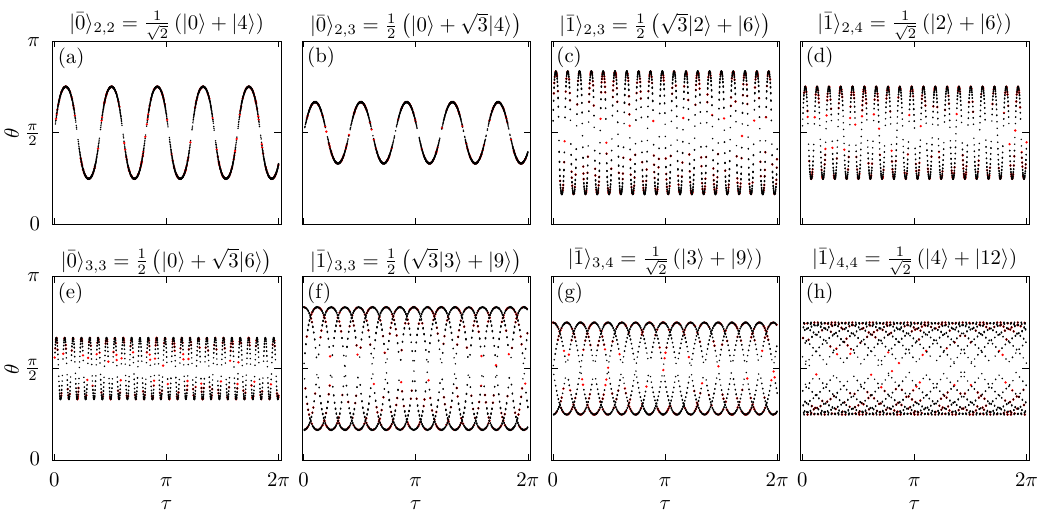}
    \vspace{-4ex}
    \caption{Combinations of $\{\tau,\, \theta\}$ for which the fidelity $F_{2g}(\tau,\,\theta)$ of the postselected oscillator state—obtained by measuring the qubit in the ground state $\ket{g}$ and given in Eq.~\eqref{eqn_fid_two_fock_ket_g}—matches each of the eight target logical binomial states with unit fidelity are shown in panels (a)–(h), respectively. The parameters $\tau$ and $\theta$ were sampled uniformly over $0 \leqslant \tau \leqslant 2\pi$ and $0 \leqslant \theta \leqslant \pi$, using 1001 and 501 points, respectively. All displayed points satisfy a numerical fidelity threshold of $|F_{2g}-1|\leqslant 10^{-4}$, with red points marking those that additionally meet the stricter threshold $|F_{2g}-1|\leqslant 10^{-6}$.}
    \label{fig:two_sup_ket_g}
\end{figure*}
%
%%%%%%%%%%%%%%%%%%%%%%%%%%%%%%%%%%%%%%%%%%%%%%%%%%%%%%%%%%%%%%%%%%%%%%%%%%%%%%%%%%%%%%%%%%%%%%%%%%%%%%%%%%%
\section{Superposition of two Fock states}
\label{sec_two_fock}
It can be readily verified from Eq.~\eqref{eqn_bonom_code_states} that there are eight logical states in total, each being a superposition of only two Fock states, given by
\begin{subequations}
\label{eqn_sup_two_fock_states}
 \begin{align}
  \ket{\bar{0}}_{2,2} &=  \tfrac{1}{\sqrt{2}} \left(\ket{0} + \ket{4}\right), \\
  \ket{\bar{0}}_{2,3} &=  \tfrac{1}{2} \left(\ket{0} + \sqrt{3}\ket{4}\right), \\
  \ket{\bar{1}}_{2,3} &=  \tfrac{1}{2} \left(\sqrt{3}\ket{2} + \ket{6}\right), \\
 \ket{\bar{1}}_{2,4} &=  \tfrac{1}{\sqrt{2}} \left(\ket{2} + \ket{6}\right),\\
  \ket{\bar{0}}_{3,3} &=  \tfrac{1}{2} \left(\ket{0} + \sqrt{3}\ket{6}\right), \\
  \ket{\bar{1}}_{3,3} &=  \tfrac{1}{2} \left(\sqrt{3}\ket{3} + \ket{9}\right),\\
  \ket{\bar{1}}_{3,4} &=  \tfrac{1}{\sqrt{2}} \left(\ket{3} + \ket{9}\right), \\
  \ket{\bar{1}}_{4,4} &=  \tfrac{1}{\sqrt{2}} \left(\ket{4} + \ket{12}\right).
\end{align}
\end{subequations}
In the following, we exploit the joint state $\ket{\Psi(\tau)}$ in Eq.~\eqref{eqn_psit_2fock} to generate all eight superpositions states. Our main protocol is probabilistic, relying on projective measurement of the qubit to herald the target oscillator state. We focus on this approach due to its ability to achieve higher fidelities and greater flexibility in parameter selection. For completeness, we also examine a deterministic generation protocol that avoids qubit measurement by tracing over the qubit degrees of freedom. The details of this alternative method, along with its limitations and comparative performance, are discussed in Sec.~\ref{sec_deterministic}.

Upon projective measurement of the qubit, if it is found in the ground state $\ket{g}$, the corresponding unnormalized oscillator state is given by $\ket{\psi_o(\tau)}_{2g} = x_1(\tau)\ket{n_1} + x_3(\tau)\ket{n_1+m}$, as follows directly from Eq.~\eqref{eqn_psit_2fock}.  It is important to note that obtaining the eight superposition states in Eq.~\eqref{eqn_sup_two_fock_states} from $\ket{\psi_o(\tau)}_{2g}$ requires $m > n_1$. Therefore, we have
\begin{equation}
	\ket{\psi_o(\tau)}_{2g} = \mathcal{N}_{2g} \left\{\cos\theta \ket{n_1} - \sin\theta\sin(a_{n_1,m}\tau)\ket{n_1+m}\right\},
	\label{eqn_osc_two_fock_ket_g}
\end{equation}
where $a_{n_1,m}=\sqrt{(n_1+m)!/n_1!}$ and the normalization  $\mathcal{N}_{2g} = \left\{\cos^2\theta+\sin^2\theta\sin^2(a_{n_1,m}\tau)\right\}^{-1/2}$. 
%$\mathcal{N} = 1/\sqrt{\cos^2\theta+\sin^2\theta\sin^2(a_{n_1,m}\tau)}$. 
%.
%  It is important to note that this state cannot be used to generate any of the desired binomial superposition states, as it requires $m$ to be much greater than $n_1$, which is not feasible.

Writing the eight target states in the form $c_0\ket{n_1} + c_1\ket{n_1+m}$ (where $|c_0|^2+|c_1|^2=1$), we can express the fidelity in the generic compact form
\begin{equation}
	F_{2g}(\tau,\,\theta) = \frac{\left| c_0\cos\theta - c_1 \sin\theta\sin(a_{n_1,m}\tau)\right|^2}{\cos^2\theta+\sin^2\theta\sin^2(a_{n_1,m}\tau)}.
	\label{eqn_fid_two_fock_ket_g}
\end{equation}
\begin{table}[ht!]
	\centering
	\caption{Counts of parameter pairs $(\tau,\theta)$ for which the fidelity $F_{2g}$ (see Eq.~\eqref{eqn_fid_two_fock_ket_g}) between the engineered two–Fock-component state in Eq.~\eqref{eqn_osc_two_fock_ket_g} and the corresponding eight target codeword satisfies $|F_{2g}-1|\le 10^{-4}$ or $10^{-6}$.  The parameters $\tau$ and $\theta$ were sampled uniformly over  $0 \leqslant \tau \leqslant 2\pi$  and  $0 \leqslant \theta \leqslant \pi$, using 1001 and 501 points, respectively. Each entry reports the number of grid points in the scanned $(\tau,\theta)$ domain for which the fidelity falls within the specified tolerance.}
	\label{tab:fidelity_counts_ket_g}
	\begin{ruledtabular}
		\begin{tabular}{lcc|lcc}
                State & $N_{10^{-4}}$ & $N_{10^{-6}}$ &State & $N_{10^{-4}}$ & $N_{10^{-6}}$ \\
                \hline
                $\ket{\bar{0}}_{2,2}$  & 2545 & 271  &
                $\ket{\bar{0}}_{3,3}$  & 2217 & 217 \\
                $\ket{\bar{0}}_{2,3}$  & 2233 & 219  &
                $\ket{\bar{1}}_{3,3}$  & 3081 & 291 \\
                $\ket{\bar{1}}_{2,3}$  & 3094 & 298  &
                $\ket{\bar{1}}_{3,4}$  & 2520 & 270 \\
                $\ket{\bar{1}}_{2,4}$  & 2522 & 271  &
                $\ket{\bar{1}}_{4,4}$  & 2504 & 260 \\
                \end{tabular}
	\end{ruledtabular}
\end{table}
By judiciously choosing the pair ($n_1, m$) one selects which member of the two-Fock–component binomial code family is to be realized. Once the values of $n_1$, and  $m$ are fixed, the remaining task is to optimize the control parameters $\theta$ and $\tau$, which determine the relative weights of the two Fock components, to achieve fidelities arbitrary close to unity.
Although the states $\ket{\psi_o(\tau)}_{2g}$ in Eq.~\eqref{eqn_osc_two_fock_ket_g} span the relevant two-dimensional Fock subspace, identifying the corresponding control parameters generally requires a numerical search. This is due to the fact that the effective coupling strength $a_{n_1,m}$ is typically irrational, preventing a closed-form analytical solution for arbitrary target amplitude ratios. We therefore perform a numerical optimization numerically by scanning $\theta \in [0, \pi]$, and $\tau \in [0, 2\pi]$ with a resolution of 501 points for $\theta$ and 1001 points for $\tau$. Numerous parameter combinations yield fidelities equal to unity for all eight states, , highlighting the robustness and flexibility of the protocol..
%With judicious choices of $n_1$ and $m$, and by appropriately tuning the parameters $\theta$ and $\tau$, all the eight two-Fock component binomial code states can be systematically synthesized. 
%As anticipated, increasing the numerical precision reduces the number of viable solutions. Under a stricter tolerance of $10^{-6}$, the number of matches drops down significantly, marked as red points in the plots.

\begin{figure*}[ht]
    \centering
    \includegraphics[width=1.0\linewidth]{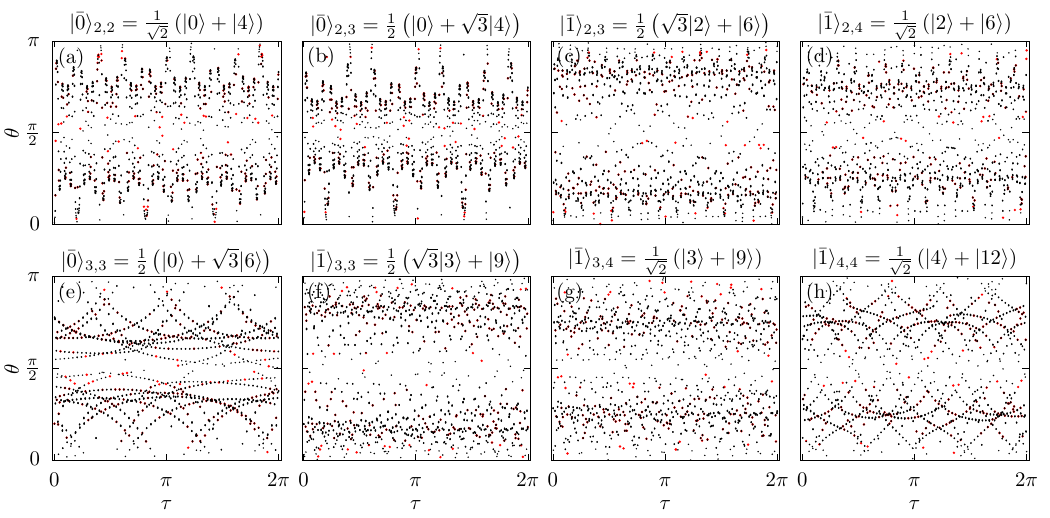}
    \vspace{-4ex}
    \caption{Combinations of $\{\tau,\, \theta\}$ for which the fidelity $F_{2e}(\tau,\,\theta)$ of the postselected oscillator state—obtained by measuring the qubit in the ground state $\ket{e}$ and given in Eq.~\eqref{eqn_fid_two_fock_ket_e}—matches each of the eight target logical binomial states with unit fidelity are shown in panels (a)–(h), respectively.. All sampling details and fidelity thresholds are identical to those used in Fig.~\ref{fig:two_sup_ket_g}.}
    \label{fig:two_sup}
\end{figure*}

These optimal combinations are visualized in Fig.~\ref{fig:two_sup_ket_g}, where each scatter point corresponds to a distinct choice of $(\theta, \, \tau)$ achieving unit fidelity under the numerical tolerance of $10^{-4}$.
As anticipated, tightening the precision requirement reduces the number of admissible solutions: under a stricter tolerance of $10^{-6}$, only the most robust matches survive, shown as red points in the plots. The details on the total count of the optimal parameter pairs are listed in Table~\ref{tab:fidelity_counts_ket_g} for all eight cases.

On the other hand, if the qubit is in the excited state $\ket{e}$ after measurement, the resulting oscillator state has the form
\begin{align}
    \ket{\psi_o(\tau)}_{2e} &= \mathcal{N}_{2e}\big\{\sin\theta\, \cos \left(a_{n_1,m}\, \tau\right)\ket{n_1}  \nonumber\\
    &\quad\quad+ \cos\theta\, \sin\left(b_{n_1,m}  \, \tau\right)\ket{n_1-m}\big\},
	\label{eqn_osc_two_fock_ket_e}
\end{align}
where $b_{n_1,m}=\sqrt{n_1!/(n_1-m)!}$ and the normalization $\mathcal{N}_{2e}=\left\{\cos^2\theta \sin^2 \left(a_{n_1,m}\, \tau\right) + \sin^2\theta \cos^2\left( b_{n_1,m} \, \tau\right) \right\}^{-\frac{1}{2}}$.

Once again, with judicious choices of $n_1$ and $m$ ($n_1\geqslant m$), and by appropriately tuning the control parameters $\theta$ and $\tau$, all eight target two-Fock superposition states can be systematically synthesized. As before, although these parameters provide full control over the two-dimensional Fock subspace, identifying those that realize a given target state generally requires a numerical optimization due to the irrational effective coupling strengths $a_{n_1,m}$ and $b_{n_1,m}$. Writing the target states in the form $c_{0}\ket{n_1-m} + c_1\ket{n_1}$, the fidelity assumes the compact expression
\begin{equation}
	F_{2e}(\tau,\theta) = \frac{\left| c_{0}\cos\theta \sin(b_{n_1,m}\tau) + c_1\sin\theta\cos(a_{n_1,m}\tau)\right|^2}{\cos^2\theta\sin^2(b_{n_1,m}\tau)+\sin^2\theta\cos^2(a_{n_1,m}\tau)}.
	\label{eqn_fid_two_fock_ket_e}
\end{equation}
%All combinations of $\theta$ and $\tau$ for which the logical states $\ket{\bar{0}}_{2,2}$ and $\ket{\bar{1}}_{2,2}$ are realized with unit fidelity are depicted in Fig.~\ref{fig:two_sup}(e,f). 
Similar to the previous case, we find multiple parameter combinations that achieve all the target binomial code states with unit fidelity, as illustrated in Fig.~\ref{fig:two_sup}. The corresponding counts of optimal parameter choices are detailed in Table~\ref{tab:fidelity_counts_ket_e}. 

It is straightforward to verify that, by following a similar procedure, arbitrary superpositions of two Fock states can be realized (see Appendix~\ref{sec_two_fock_span} for further details). In the following section, we extend this approach to generate superpositions of three Fock states.

\begin{table}[ht]
	\centering
	\caption{Counts of parameter pairs $(\tau,\theta)$ for which the fidelity $F_{2e}$ (see Eq.~\eqref{eqn_fid_two_fock_ket_e}) satisfies $|F_{2e}-1|\le 10^{-4}$ or $10^{-6}$.  The remaining details are the same as in Table.~\ref{tab:fidelity_counts_ket_g}.}
	\label{tab:fidelity_counts_ket_e}
	\begin{ruledtabular}
		\begin{tabular}{lcc|lcc}
                State & $N_{10^{-4}}$ & $N_{10^{-6}}$ &State & $N_{10^{-4}}$ & $N_{10^{-6}}$ \\
                \hline
                $\ket{\bar{0}}_{2,2}$  & 2347 & 267  &
                $\ket{\bar{0}}_{3,3}$  & 2508 & 265 \\
                $\ket{\bar{0}}_{2,3}$  & 2514 & 255  &
                $\ket{\bar{1}}_{3,3}$  & 2381 & 228 \\
                $\ket{\bar{1}}_{2,3}$  & 2502 & 255  &
                $\ket{\bar{1}}_{3,4}$  & 2521 & 248 \\
                $\ket{\bar{1}}_{2,4}$  & 2375 & 242  &
                $\ket{\bar{1}}_{4,4}$  & 2366 & 253 \\
                \end{tabular}
	\end{ruledtabular}
\end{table}

%%%%%%%%%%%%%%%%%%%%%%%%%%%%%%%%%%%%%%%%%%%%%%%%%%%%%%%%%%%%%%%%%%%%%%%%%%%%%%%%%%%%%%%%%%%%%%%%%%%%%%%%%%%

\section{Superposition~of~three~Fock~states}
\label{sec_three_fock}
 %A straightforward inspection of Eq.~\eqref{eqn_bonom_code_states} reveals that there are sixteen logical states, each written as a superposition of three Fock states. 
 A straightforward inspection of Eq.~\eqref{eqn_bonom_code_states} reveals that there exist sixteen logical states in total, each being a superposition of only three Fock states. Out of which the seven logical $\ket{\bar{0}}$ states (3 for $K=4$ and 4 for $K=5$) are given by
\begin{subequations}
\begin{align}
\ket{\bar{0}}_{N,4} &= \frac{1}{2\sqrt{2}} \left( \ket{0} + \sqrt{6}\ket{2N} + \ket{4N} \right), \\
\ket{\bar{0}}_{N,5} &= \frac{1}{4} \left( \ket{0} + \sqrt{10}\ket{2N} + \sqrt{5}\ket{4N} \right).
\end{align}
\label{eqn_logical_0_states}
 \end{subequations}
 \begin{figure*}
    \centering
    \includegraphics[width=1.0\linewidth, height=8cm]{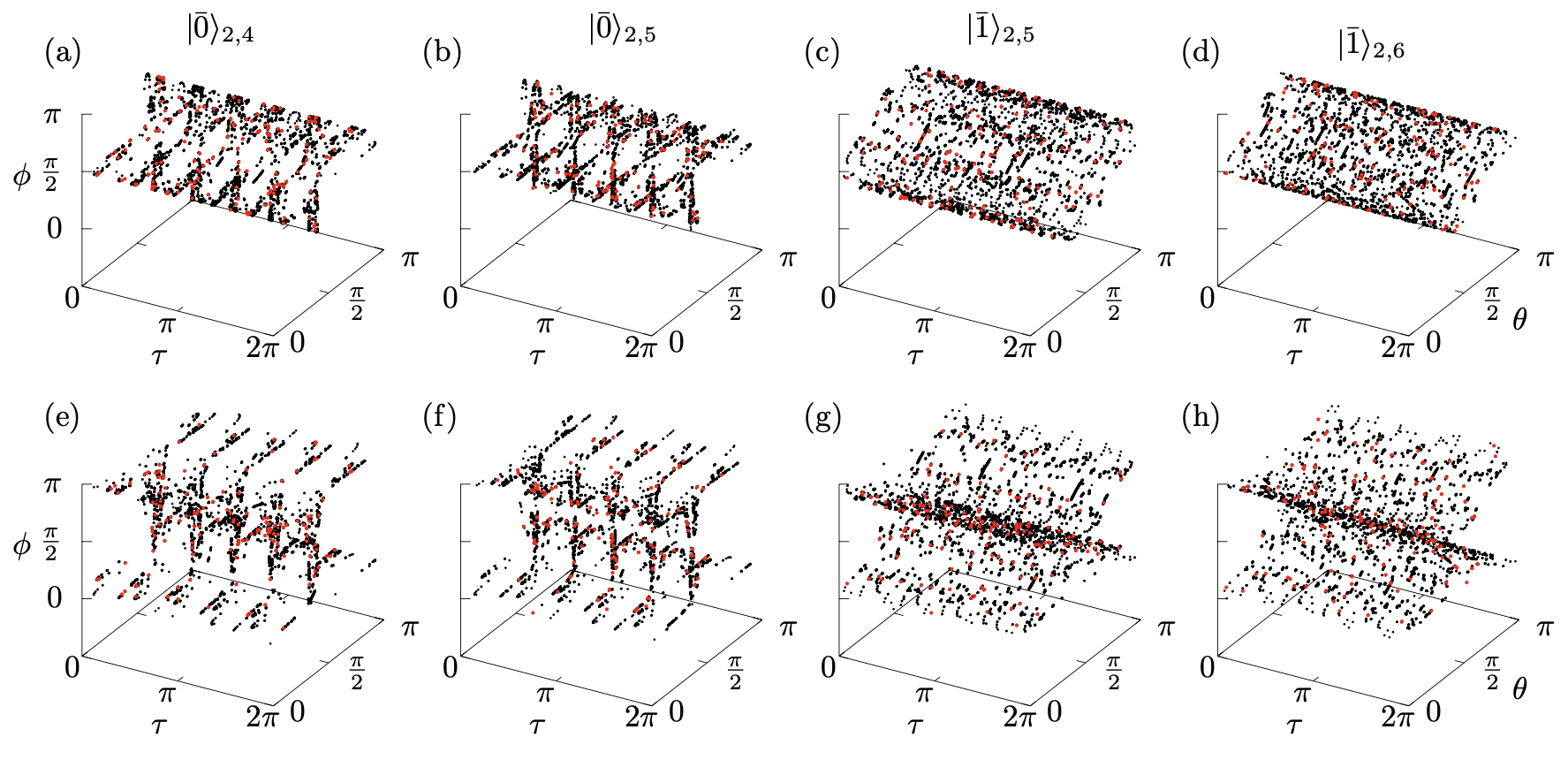}
    \vspace{-4ex}
    \caption{Combinations of $\{\tau, \theta, \phi\}$ for which the fidelity ($F_{3g_\pm}$ in Eq.~\eqref{eqn_fid_three_fock_ket_3g_pm}) of the postselected oscillator state in Eq.~\eqref{eqn_psit_3g_both} reaches unity (within numerical precision) with each of the four three-component binomial targets $\ket{\bar{0}}_{2,4}= \frac{1}{2\sqrt{2}} \left( \ket{0} + \sqrt{6}\ket{4} + \ket{8} \right)$, $\ket{\bar{0}}_{2,5}=\frac{1}{4} \left( \ket{0} + \sqrt{10}\ket{4} + \sqrt{5}\ket{8} \right)$, $\ket{\bar{1}}_{2,5}=\frac{1}{4}\left(\sqrt{5}\ket{2} + \sqrt{10}\ket{6} + \ket{10}\right)$, and $\ket{\bar{1}}_{2,6}=\frac{1}{4 \sqrt{2}}\left(\sqrt{6}\ket{2} + \sqrt{20}\ket{6} + \sqrt{6}\ket{10}\right)$ are shown. The top row corresponds to the case $n_2 = n_1 + m$ ($F_{3g_+}$), while the bottom row corresponds to $n_2 = n_1 - m$ ($F_{3g_-}$). The parameters were sampled uniformly over $0 \le \tau \le 2\pi$ (1001 points) and $0 \le \theta,\phi \le \pi$ (501 points each). As in Figs.~\ref{fig:two_sup_ket_g} and~\ref{fig:two_sup}, all plotted points satisfy $|F - 1| \le 10^{-4}$, with red points indicating those additionally meeting the stricter threshold $|F - 1| \le 10^{-6}$.}
    \label{fig:three_sup_ket_g}
\end{figure*}
The remaining nine logical $\ket{\bar{1}}$ states (4 for $K=5$ and 5 for $K=6$) are given by
\begin{subequations}
\begin{align}
\ket{\bar{1}}_{N,5} &= \frac{1}{4}\left(\sqrt{5}\ket{N} + \sqrt{10}\ket{3N} + \ket{5N}\right), \\
\ket{\bar{1}}_{N,6} &= \frac{1}{4 \sqrt{2}}\left(\sqrt{6}\ket{N} + \sqrt{20}\ket{3N} + \sqrt{6}\ket{5N}\right).
\end{align}
\label{eqn_logical_1_states}
 \end{subequations}
In all cases, $N=2,3,...,K$.

%The complete set of these sixteen states is provided in Appendix~\ref{sec_app_state_vector} (see Eqs.~\eqref{eqn_logical_0_states} and \eqref{eqn_logical_1_states}).
In the preceding section, we demonstrated that, by employing the MPJC Hamiltonian and appropriately selecting the initial configurations, arbitrary superpositions of two Fock states can be engineered. In what follows, we extend this method to synthesize superpositions of three Fock states with high fidelity, by using an initial oscillator state that is itself an arbitrary superposition of two Fock states, along with an initial arbitrary superposition of the qubit basis states. 
%\subsection{Time-evolved state vector of MPJC Hamiltonian}
That is, we assume the initial state for the bipartite system to be
%that both the qubit and the oscillator are initially prepared in superposition states  $\left(\cos\theta\ket{g}+\sin\theta\ket{e}\right)$ and $\left(\cos\phi\ket{n_1}+\sin\phi\ket{n_2}\right)$ such that 
\begin{align}
    \ket{\Psi(0)} &= \left(\cos\theta\ket{g}+\sin\theta\ket{e}\right)\otimes\left(\cos\phi\ket{n_1}+\sin\phi\ket{n_2}\right).% \nonumber\\
   % &=\cos\theta\cos\phi\ket{g, n_1}+\sin\theta\cos\phi\ket{e, n_1} \nonumber\\
    %&\quad+\cos\theta\sin\phi\ket{g, n_2}+\sin\theta\sin\phi\ket{e, n_2}.
\end{align}
% \begin{figure*}
%     \centering
%     \includegraphics[width=0.245\linewidth]{fidelity_F1.pdf}
%     \includegraphics[width=0.245\linewidth]{fidelity_F2.pdf}
%     \includegraphics[width=0.245\linewidth]{fidelity_F3.pdf}
%     \includegraphics[width=0.245\linewidth]{fidelity_F4.pdf}
%     \vspace{-4ex}
%     \caption{Combinations of $\{\tau, \theta, \phi\}$ for which the fidelities of the time-evolved oscillator states in Eq.~\eqref{eqn_psi_t_e_3_cor} match the target logical states $\ket{\bar{0}}_{3,1}$, $\ket{\bar{0}}_{4,1}$, $\ket{\bar{1}}_{4,1}$, and $\ket{\bar{1}}_{5,1}$ with unit fidelity, up to a numerical precision of $10^{-4}$ are shown in panels (a) to (d), respectively. A total of 1001 (501) values of $\tau$ ($\theta$ and $\phi$) were sampled numerically over the ranges $0 \leqslant \tau \leqslant 2\pi$ ($0 \leqslant \theta,\phi \leqslant \pi$).}
%     \label{fig:three_sup}
% \end{figure*}
%%%%%%%%%%%%%%%%%%%%%%%%%%%%%%%%%%%%%%%%%%%%%%%%%%%%%%%%%%%%%%%%%%%%%%%%%%%%%%%%%%

It is worth noting that, although such an initial state formally departs from our first assumption, we showed in the previous section that arbitrary two-Fock superpositions can still be generated using the MPJC Hamiltonian. Building on this, three-Fock-state superpositions can be realized in a two-step protocol: first prepare a two-Fock resource state, then extend it to the desired three-Fock superposition.
It can be straightforwardly shown that the corresponding time-evolved state can be written compactly as a superposition of two independent two-Fock dynamical blocks. Explicitly, one finds
\begin{align}
\ket{\Psi(\tau)} &= \sum_{i=1}^{4} x_i(\tau)\,\ket{\chi_i(n_1)} + \sum_{i=1}^{4} y_i(\tau)\,\ket{\chi_i(n_2)},
\label{eqn_psit_3fock}
\end{align}
where $\ket{\chi_i(n)} \in {\ket{g,n},\ket{e,n},\ket{g,n+m},\ket{e,n-m}}$.
The time-dependent amplitudes factorize as
\begin{subequations}
\begin{align}
x_i(\tau) &= \cos\phi\, x_i^{(2)}(n_1,m,\theta, \tau),\\
y_i(\tau) &= \sin\phi\, x_i^{(2)}(n_2,m,\theta, \tau),
\end{align}
\label{eqn_xyt_3fock_short}
\end{subequations}
\begin{table}[ht]
\centering
\caption{The coefficients $C_{lg}^{(\pm)}$ and $C_{le}^{(\pm)}$ for the three–Fock-component states. Note that $a_{n_1,m}=\sqrt{(n_1+m)!/n_1!}$, $b_{n_1,m}=\sqrt{n_1!/(n_1-m)!}$, and $c_{n_1,m} = \sqrt{(n_1+2m)!/(n_1+m)!}$.}
\label{tab:coeff_three_fock}
\begin{ruledtabular}
\begin{tabular}{l l}
$C_{0g}^{(+)}$ 
    & $\cos\theta\cos\phi$
    \\
$C_{1g}^{(+)}$ 
    & $\cos\theta\sin\phi\,\cos(a_{n_1,m}\tau) - \sin\theta\cos\phi\,\sin(a_{n_1,m}\tau)$
    \\
$C_{2g}^{(+)}$
    & $-\sin\theta\sin\phi\,\sin(c_{n_1,m}\tau)$
    \\
\hline
$C_{0g}^{(-)}$
    & $\cos\theta\sin\phi$
    \\
$C_{1g}^{(-)}$
    & $\cos\theta\cos\phi\,\cos(b_{n_1,m}\tau) - \sin\theta\sin\phi\,\sin(b_{n_1,m}\tau)$
    \\
$C_{2g}^{(-)}$
    & $-\sin\theta\cos\phi\,\sin(a_{n_1,m}\tau)$
    \\
\hline
$C_{0e}^{(+)}$
    & $\cos\theta\cos\phi\,\sin(b_{n_1,m}\tau)$
    \\
$C_{1e}^{(+)}$
    & $\sin\theta\cos\phi\,\cos(a_{n_1,m}\tau) + \cos\theta\sin\phi\,\sin(a_{n_1,m}\tau)$
    \\
$C_{2e}^{(+)}$
    & $\sin\theta\sin\phi\,\cos(c_{n_1,m}\tau)$
    \\
\hline
$C_{0e}^{(-)}$
    & $\cos\theta\sin\phi$
    \\
$C_{1e}^{(-)}$
    & $\cos\theta\cos\phi\,\sin(b_{n_1,m}\tau) + \sin\theta\sin\phi\,\cos(b_{n_1,m}\tau)$
    \\
$C_{2e}^{(-)}$
    & $\sin\theta\cos\phi\,\cos(a_{n_1,m}\tau)$
    \\
\end{tabular}
\end{ruledtabular}
\end{table}
where $x_i^{(2)}$ denotes the coefficients of the two-Fock solution given previously in Eq.~\eqref{eqn_xt_2fock}. Thus the three-Fock evolution decomposes naturally into two independent excitation manifolds, weighted by $\cos\phi$ and $\sin\phi$, respectively. For $n_1<m$, $x_1(\tau) = \cos\theta \cos\phi, \quad  x_4(\tau) = 0$ and for $n_2<m$, $y_1(\tau) = \cos\theta\sin\phi, \quad  y_4(\tau) = 0$ (see Appendix~\ref{sec_app_state_vector} for details). The two-Fock solution is recovered simply by setting $\phi=0$.

Now, suppose a projective measurement is performed on the qubit, and the outcome is the ground state $\ket{g}$. The oscillator is then projected onto a four-Fock superposition state given by
\begin{align}
    \ket{\psi_o(\tau)}_{4g} &= \mathcal{N}_{4g}\big\{x_1(\tau)\ket{n_1} + x_3(\tau)\ket{n_1+m} \nonumber\\
   				 &\qquad + y_1(\tau)\ket{n_2} + y_3(\tau)\ket{n_2+m}  \big\},
    \label{eqn_psi_t_g_3}
\end{align}
where $\mathcal{N}_{4g}$ is the normalization factor.
%where $\mathcal{N}=\left\{|x_1|^2+|x_3|^2+|y_1|^2+|y_3|^2\right\}^{-1/2}$.

The four-Fock component superposition state in Eq.~\eqref{eqn_psi_t_g_3} can be reduced to a three distinct components  only when the pairs $\{n_1,n_1+m\}$ and $\{n_2,n_2+m\}$ overlap on exactly one element.  
Given that all Fock numbers are nonnegative integers, two possibilities arise. These are (i) $n_2 = n_1 + m$, and  (ii) $n_2 = n_1 - m$ (with $n_1\ge m$).

%\paragraph*{Case (i): $n_2=n_1+m$.} 
In the case $n_2 = n_1 + m$, the accessible Fock ladder is
${k_0, k_1, k_2} = {n_1\,, n_1+m\,, n_1+2m}$. Note that for all the sixteen logical states, this requires $n_1 < m$ (see Eqs.~\eqref{eqn_logical_0_states} and \eqref{eqn_logical_1_states}).
For the alternative choice $n_2 = n_1 - m$ with $n_1 \ge m$, the ladder becomes
${k_0, k_1, k_2} = {n_1-m,\, n_1\,, n_1+m}$.
In both cases, the oscillator state derived from Eq.~\eqref{eqn_psi_t_g_3} can be written in the unified form
\begin{align}
	\label{eqn_psit_3g_both}
	\ket{\psi_o(\tau)}_{3g_{\pm}} = \mathcal{N}_{3g_\pm} \sum_{l=0}^2 C_{lg}^{(\pm)}(\tau, \theta,\phi)\, \ket{k_l},
\end{align}
where the choice of ${k_0,k_1,k_2}$ is determined by whether $n_2 = n_1 + m$ (`$+$') or $n_2 = n_1 - m$ (`$-$'). The exact expressions of the coefficients $C_{lg}^{(\pm)}$ can be found in Table~\ref{tab:coeff_three_fock}.  The normalization factor $\mathcal{N}_{3g_\pm}  = \left\{|C_{0g}^{(\pm)}|^2+|C_{1g}^{(\pm)}|^2+|C_{2g}^{(\pm)}|^2\right\}^{-1/2}$.

Writing the sixteen target states in the form $\sum_{l}c_l\ket{k_l}$ for both the cases, the fidelity for both constructions can be succinctly expressed in the form
\begin{equation}
%	F_{3g_{\pm}}(\tau, \theta,\phi) = \left| \sum\limits_{k=0}^2 c_0 C_{kg}^{(\pm)}\right|^2/ \sum\limits_{k=0}^2 \left|C_{kg}^{(\pm)} \right|^2  .
	F_{3g_{\pm}}(\tau, \theta,\phi) = \frac{\left| \sum\limits_{l=0}^2 c_l\, C_{lg}^{(\pm)}(\tau, \theta,\phi)\right|^2}{ \sum\limits_{l=0}^2 \left|C_{lg}^{(\pm)}(\tau, \theta,\phi) \right|^2  }.
%	F_{3g_{\pm}}(\tau, \theta,\phi) = \frac{\left| c_0 C_{0g}^{(\pm)} + c_1 C_{1g}^{(\pm)}+ c_2 C_{2g}^{(\pm)} \right|^2}{\left|C_{0g}^{(\pm)} \right|^2 +  \left|C_{1g}^{(\pm)}\right|^2 + \left|C_{2g}^{(\pm)}\right|^2 }.
	\label{eqn_fid_three_fock_ket_3g_pm}
\end{equation}

In what follows, we illustrate the procedure using four representative examples: the binomial code states  $\ket{\bar{0}}_{2,4}$, $\ket{\bar{0}}_{2,5}$, $\ket{\bar{1}}_{2,5}$, and $\ket{\bar{1}}_{2,6}$. The first pair are superpositions of $\ket{0}$, $\ket{4}$, and $\ket{8}$, while the second pair are superpositions of $\ket{2}$, $\ket{6}$, and $\ket{10}$. For the choice $n_2 = n_1 + m$, these pairs correspond to $n_1 = 0$ and $n_1 = 2$, respectively. For the alternative choice $n_2 = n_1 - m$, they correspond instead to $n_1 = 4$ and $n_1 = 6$. In both constructions, the spacing is fixed at $m = 4$.

\begin{figure*}
    \centering
    \includegraphics[width=1.0\linewidth, height=8cm]{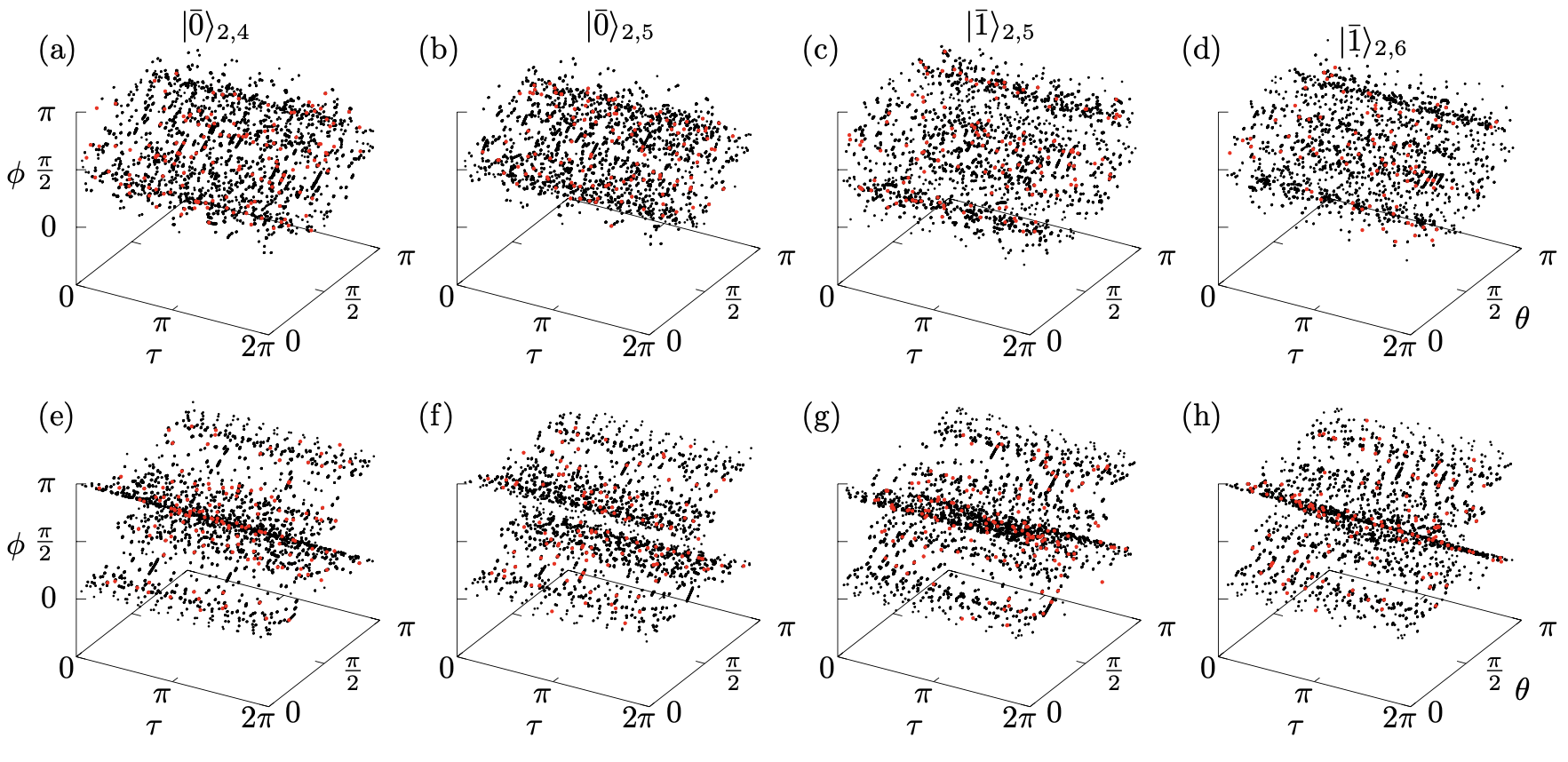}
    \vspace{-4ex}
    \caption{Combinations of $\{\tau, \theta, \phi\}$ for which the fidelity ($F_{2e_\pm}$ in Eq.~\eqref{eqn_fid_three_fock_ket_3e_pm}) of the postselected oscillator state in Eq.~\eqref{eqn_psit_3e_both} reaches unity (within numerical precision) with each of the four three-component binomial targets $\ket{\bar{0}}_{2,4}$, $\ket{\bar{0}}_{2,5}$, $\ket{\bar{1}}_{2,5}$, and $\ket{\bar{1}}_{2,6}$ are shown. As in Fig.~\ref{fig:three_sup_ket_g}, the top row corresponds to the case $n_2 = n_1 + m$ ($F_{2e_+}$), while the bottom row corresponds to $n_2 = n_1 - m$ ($F_{2e_-}$. All sampling details and fidelity thresholds are identical to those used in Fig.~\ref{fig:three_sup_ket_g}.}
    \label{fig:three_sup_ket_e}
\end{figure*}

\begin{table}[ht]
	\centering
	\caption{Counts of parameter triplets $(\tau,\theta,\phi)$ for which the fidelity $F_{3g_{\pm}}$ (see Eq.~\eqref{eqn_fid_three_fock_ket_3g_pm}) between the engineered three–Fock-component state in Eq.~\eqref{eqn_psit_3g_both} and the corresponding sixteen target codeword satisfies $|F_{3g_{\pm}}-1|\le 10^{-4}$ or $10^{-6}$.  The parameters $\tau$, $\theta$, and $\phi$ were sampled uniformly over  $0 \leqslant \tau \leqslant 2\pi$, $0 \leqslant \theta \leqslant \pi$  and  $0 \leqslant \phi \leqslant \pi$, using 1001, 501 and 501 points, respectively. Each entry reports the number of grid points in the scanned $(\tau,\theta,\phi)$ domain for which the fidelity falls within the specified tolerance.}
	\label{tab:fidelity_counts_ket_g_3sup}
	\begin{ruledtabular}
		\begin{tabular}{lcc|lcc}
            \multicolumn{3}{c|}{$n_2 = n_1 + m$} & \multicolumn{3}{c}{$n_2 = n_1 - m$} \\ \hline
            State & $N_{10^{-4}}$ & $N_{10^{-6}}$ & State & $N_{10^{-4}}$ & $N_{10^{-6}}$ \\
            \hline
            $\ket{\bar{0}}_{2,4}$  & 13097 &  147 & $\ket{\bar{0}}_{2,4}$  & 13097 &  147\\
            $\ket{\bar{0}}_{2,5}$  & 12418 &  135 & $\ket{\bar{0}}_{2,5}$  & 12418 &  135\\
            $\ket{\bar{1}}_{2,5}$  & 16297 &  161 & $\ket{\bar{1}}_{2,5}$  & 16297 &  161\\
            $\ket{\bar{1}}_{2,6}$  & 13908 &  144 & $\ket{\bar{1}}_{2,6}$  & 13908 & 144 \\
        \end{tabular}
	\end{ruledtabular}
\end{table}

Similar to the two–Fock-component case, the task now reduces to optimizing the control parameters—here $\theta$, $\phi$, and $\tau$—to achieve unit fidelity for each target code state. This optimization is again performed numerically. As explained before, this numerical search is necessitated by the generally irrational nature of the effective coupling strengths, which prevents a closed-form analytical determination of the control parameters for arbitrary target states.  Specifically, we scan the parameter ranges $\theta \in [0, \pi]$, $\phi \in [0, \pi]$, and $\tau \in [0, 2\pi]$ using a grid of 501 points for  $\theta$ and $\phi$, and 1001 points for $\tau$. As in the two–Fock case, a large number of parameter triples yield unit fidelities, up to numerical precision.
%Specifically, we identify 14809 valid parameter sets for $F_1$, 13311 for $F_2$, 13601 for $F_3$, and 10752 for $F_4$, each achieving unit fidelity within a numerical precision of $10^{-4}$. 
These optimal combinations are visualized as scatter plots in Fig.~\ref{fig:three_sup_ket_g}, highlighting the substantial flexibility in selecting parameters for synthesizing the target code states. As expected, increasing the numerical precision  to $10^{-6}$ reduces the number of viable solutions (marked as red points in the plots). Further details are provided in Table~\ref{tab:fidelity_counts_ket_g_3sup}.

On the other hand, after the measurement, if the qubit is found in the excited state $\ket{e}$, the oscillator gets projected onto a four-Fock superposition state given by
\begin{align}
    \ket{\psi_o(\tau)}_{4e} =  \mathcal{N}_{4e}&\big\{ x_2(\tau)\ket{n_1} + x_4(\tau)\ket{n_1-m}\nonumber\\
    							&+y_2(\tau)\ket{n_2} +y_4(\tau)\ket{n_2-m}\big\},
    \label{eqn_psi_t_e_3}
\end{align}
where $\mathcal{N}_{4e}$ is the normalization factor.
 
%Following similar analysis, we find the equivalent generic expressions for the  state vector and the fidelity. The accessible Fock ladders for $n_2=n_1+m$ are $\ket{n_1-m} $, $\ket{n_1}$, and $\ket{n_1+m} $ whereas for the case these are $n_2=n_1-2m$ are $\ket{n_1-m} $, $\ket{n_1}$, and $\ket{n_1} $. The state vector is therefor expressed as
Following the same analysis as for the ground-state projection, the accessible Fock ladders again fall into two cases. For $n_2=n_1+m$, the relevant oscillator subspace is ${k_0,k_1,k_2}={n_1-m\,,n_1\,,n_1+m}$; for $n_2=n_1-m$, it is ${k_0,k_1,k_2}={n_1-2m\,,n_1-m,\,n_1}$.
In analogy with the earlier analysis, the postselected state arising in both cases assumes the unified form
\begin{align}
	\label{eqn_psit_3e_both}
	\ket{\psi_o(\tau)}_{3e_{\pm}} = \mathcal{N}_{3e_\pm} \sum_{l=0}^2 C_{le}^{(\pm)}(\tau, \theta,\phi)\, \ket{k_l},
\end{align}
where the exact expressions of the coefficients are once again tabulated in Table~\ref{tab:coeff_three_fock}.
%$C_{0e}^{(+)} = \cos\theta\cos\phi\sin(b_{n_1,m}\tau)$, 
%$C_{1e}^{(+)} = \sin\theta\cos\phi\,\cos(a_{n_1,m}\tau) + \cos\theta\sin\phi\,\sin(a_{n_1,m}\tau) $, 
%$C_{2e}^{(+)} = \sin\theta\sin\phi\,\cos(c_{n_1,m}\tau)$,
% $C_{0e}^{(-)} = \cos\theta\sin\phi$ (as $n_1< 2m$ for all sixteen target binomial states of this type),
% $C_{1e}^{(-)} =  \cos\theta\cos\phi\,\sin(b_{n_1,m}\tau) +\sin\theta\sin\phi\,\cos(b_{n_1,m}\tau)$,
% $C_{2e}^{(-)} =  \sin\theta\cos\phi \cos(a_{a_1,m}\tau)$,
The normalization $\mathcal{N}_{3e_\pm}  = \left\{|C_{0e}^{(\pm)}|^2+|C_{1e}^{(\pm)}|^2+|C_{2e}^{(\pm)}|^2\right\}^{-1/2}$.

 %$C_{0}^{(-)} = y_1 = \cos\theta\sin\phi\, \cos(b_{d_1,m}\tau)$ for $n_1\geqslant 2m$ and 
%\begin{align}
%    \label{eqn_psit_3e_+}
%	\ket{\psi_o(\tau)}_{3g_{\pm}} = \mathcal{N}_{\pm} \sum_{l=0}^2 C_{lg}^{(\pm)}(\tau, \theta,\phi)\, \ket{k_l},
%    \ket{\psi_o(\tau)}_{3e_+} &= \mathcal{N}_{+}\Big\{C_{0e}^{(+)} \ket{n_1-m} + C_{1e}^{(+)}\ket{n_1} \nonumber \\
%                 &\qquad\quad + C_{2e}^{(+)} \ket{n_1+m}\Big\},\\
%    \ket{\psi_o(\tau)}_{3g_-} &= \mathcal{N}_{-}\Big\{C_{0e}^{(-)}\ket{n_1-2m} + C_{1e}^{(-)}\ket{n_1-m} \nonumber \\
%                 &\qquad\quad + C_{2e}^{(-)} \ket{n_1}\Big\}.
%    \label{eqn_psit_3e_-}
%\end{align}

As before, by expressing each of the sixteen target states in the generic form $\sum_{l} c_l \ket{k_l}$, the fidelity for both constructions can be written compactly as
\begin{equation}
	F_{3e_{\pm}}(\tau, \theta,\phi) = \frac{\left| \sum\limits_{l=0}^2 c_l C_{le}^{(\pm)}(\tau, \theta,\phi)\right|^2}{ \sum\limits_{k=0}^2 \left|C_{le}^{(\pm)}(\tau, \theta,\phi) \right|^2  }.
%	F_{3e_{\pm}}(\tau, \theta,\phi) = \frac{\left| c_0 C_{0e}^{(\pm)} + c_1 C_{1e}^{(\pm)}+ c_2 C_{2e}^{(\pm)} \right|^2}{\left|C_{0e}^{(\pm)} \right|^2 +  \left|C_{1e}^{(\pm)}\right|^2 + \left|C_{2e}^{(\pm)}\right|^2 }.
	\label{eqn_fid_three_fock_ket_3e_pm}
\end{equation}

\begin{table}[ht]
	\centering
	\caption{Counts of parameter triplets $(\tau,\theta,\phi)$ for which the fidelity $F_{3e_{\pm}}$ (see Eq.~\eqref{eqn_fid_three_fock_ket_3e_pm}) between the engineered three–Fock-component state in Eq.~\eqref{eqn_psit_3e_both} and the corresponding sixteen target codeword satisfies $|F_{3e_{\pm}}-1|\le 10^{-4}$ or $10^{-6}$.  The remaining details are identical to those in Table~\ref{tab:fidelity_counts_ket_g_3sup}.}
	\label{tab:fidelity_counts_ket_e_3sup}
	\begin{ruledtabular}
		\begin{tabular}{lcc|lcc}
            \multicolumn{3}{c|}{$n_2 = n_1 + m$} & \multicolumn{3}{c}{$n_2 = n_1 - m$} \\ \hline
            State & $N_{10^{-4}}$ & $N_{10^{-6}}$ & State & $N_{10^{-4}}$ & $N_{10^{-6}}$ \\
            \hline
            $\ket{\bar{0}}_{2,4}$  & 14809 & 159  & $\ket{\bar{0}}_{2,4}$  & 13224 &  146\\
            $\ket{\bar{0}}_{2,5}$  & 13311 &  131 & $\ket{\bar{0}}_{2,5}$  & 12496 &  133\\
            $\ket{\bar{1}}_{2,5}$  & 13601 & 139  & $\ket{\bar{1}}_{2,5}$  & 16357 &  170\\
            $\ket{\bar{1}}_{2,6}$  & 10752 & 89   & $\ket{\bar{1}}_{2,6}$  & 13587 &  150\\
        \end{tabular}
	\end{ruledtabular}
\end{table}

We again perform a numerical optimization over the control parameters $\theta$, $\phi$, and $\tau$ across the same ranges as in the previous analysis, and find that many parameter combinations yield unit fidelity. The corresponding scatter plots are shown in Fig.~\ref{fig:three_sup_ket_e}, further illustrating the substantial flexibility available when synthesizing the target code states in this setting. A summary of the fidelity counts is provided in Table~\ref{tab:fidelity_counts_ket_e_3sup}. Although the optimal parameter sets appear sparse in Figs.~\ref{fig:two_sup_ket_g}–\ref{fig:three_sup_ket_e}, their finite clustering demonstrates that the protocol tolerates small deviations in the control parameters, indicating robustness against moderate experimental imperfections in $\tau$, $\theta$, and~$\phi$.

\begin{figure*}[ht!]
    \centering
    \includegraphics[width=1.0\linewidth]{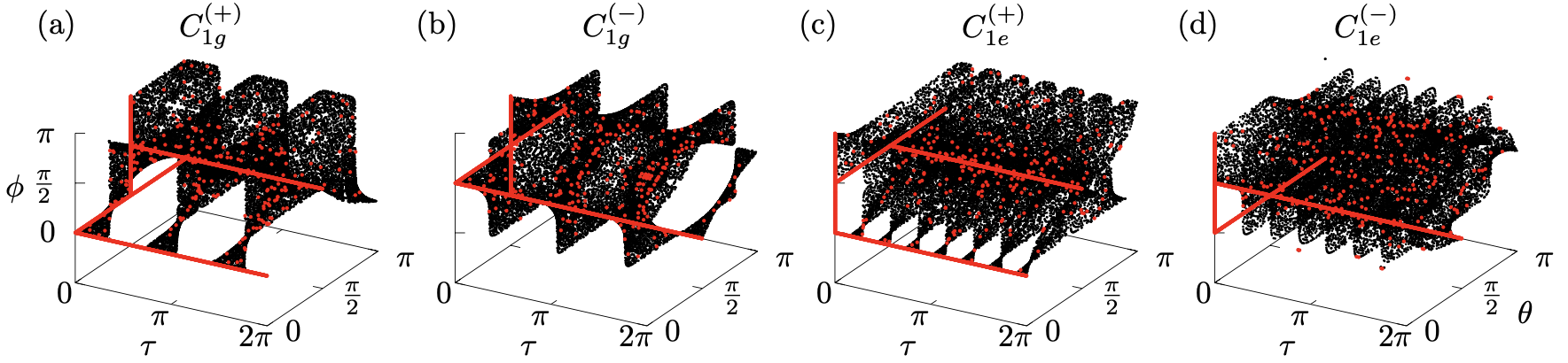}
    \vspace{-2ex}
    \caption{Combinations of $\{\tau, \theta, \phi\}$ for which the functions $C_{1g}^{(\pm)}$ and $C_{1e}^{(\pm)}$ defined in Table~\ref{tab:coeff_three_fock} vanish. Numerical sampling was performed over 1001 values of $\tau$ in the range $[0,,2\pi]$ and 501 values each for $\theta$ and $\phi$ in $[0,,\pi]$. All displayed points satisfy a numerical tolerance of $\leqslant 10^{-4}$, while red points highlight those meeting a stricter threshold of $\leqslant 10^{-6}$.}
    \label{fig:c2c3}
\end{figure*}

Before proceeding, we note that this procedure can be systematically extended to generate larger superpositions involving four or more Fock states. In principle, the four-component Fock states can be readily realized from Eqs.~\eqref{eqn_psi_t_g_3} and \eqref{eqn_psi_t_e_3} by selecting suitable values of $n_1,\, n_2,\, m$ and subsequently optimizing the control parameters. An explicit example for a four-component state is provided in Appendix~\ref{sec_app_state_vector}. Importantly, as the number of target Fock components increases further, the number of required MPJC interaction steps grows accordingly, leading to a corresponding increase in the computational overhead for numerical optimization.

%%%%%%%%%%%%%%%%%%%%%%%%%%%%%%%%%%%%%%%%%%%%%%%%%%%%%%%%%%%%%%%%%%%%%%%%%%%%%%%%%%%%%%%%%%%%%%%%%%%

\section{Reducing order of multiphoton interactions}
\label{sec_reduced_m}
Thus far, in engineering binomial code states, we have observed that increasing the number of correctable losses ($L$) necessitates a corresponding increase in the value of $m$. As a result, generating code states with large $L$ rapidly becomes practically infeasible. In this section, we address this limitation and demonstrate that the required $m$ can, in fact, be reduced by a factor of two when the code states are superpositions of two Fock states. This comes at the cost of increasing the number of times the MPJC interaction needs to be implemented.

%Recall that a single MPJC step yields superpositions of $\ket{n_1}$ and $\ket{n_1+m}$ (or $\ket{n_1}$ and $\ket{n_1-m}$) when postselecting on the qubit ground (or excited) state (see Eqs.~\eqref{eqn_osc_two_fock_ket_g} and~\eqref{eqn_osc_two_fock_ket_e}). Instead, we can use the three-Fock  superposition states $\ket{\psi_o(\tau)}_{3g{\pm}}$ and $\ket{\psi_o(\tau)}_{3e{\pm}}$ (in Eqs.~\eqref{eqn_psit_3g_both} and \eqref{eqn_psit_3e_both}, respectively), and still end up with two-Fock superposition states by eliminating the respective intermediate Fock components $C_{1g}^{(\pm)}$ or $C_{1e}^{(\pm)}$ by appropriately tuning the three control parameters $\theta$, $\phi$, and~$\tau$.  This yields superpositions of $\ket{n_1}$ and $\ket{n_1 + 2m}$ for $\ket{\psi_o(\tau)}_{3g_{+}}$, superpositions of  $\ket{n_1-m}$ and $\ket{n_1 + m}$ for $\ket{\psi_o(\tau)}_{3g_{-}}$ and $\ket{\psi_o(\tau)}_{3e_{+}}$, and superpositions of $\ket{n_1}$ and  $\ket{n_1 - 2m}$ for $\ket{\psi_o(\tau)}_{3e_{-}}$, effectively doubling the separation between Fock components in each case while keeping the MPJC interaction order fixed.

Recall that a single MPJC step produces superpositions of 
$\ket{n_1}$ and $\ket{n_1+m}$ (or $\ket{n_1}$ and $\ket{n_1-m}$) upon postselecting the qubit in the ground (or excited) state (see Eqs.~\eqref{eqn_osc_two_fock_ket_g} and~\eqref{eqn_osc_two_fock_ket_e}). Alternatively, one can employ the three-Fock superposition states  $\ket{\psi_o(\tau)}_{3g{\pm}}$ and $\ket{\psi_o(\tau)}_{3e{\pm}}$ (Eqs.~\eqref{eqn_psit_3g_both} and~\eqref{eqn_psit_3e_both}) and eliminate the intermediate Fock components $C_{1g}^{(\pm)}$ or $C_{1e}^{(\pm)}$ by appropriately tuning the three control parameters $\theta$, $\phi$, and $\tau$. This procedure yields superpositions of $\ket{n_1}$ and $\ket{n_1 + 2m}$ for $\ket{\psi_o(\tau)}_{3g_{+}}$,  superpositions of $\ket{n_1-m}$ and $\ket{n_1 + m}$ for  $\ket{\psi_o(\tau)}_{3g_{-}}$ and $\ket{\psi_o(\tau)}_{3e_{+}}$, and  superpositions of $\ket{n_1}$ and $\ket{n_1 - 2m}$ for  $\ket{\psi_o(\tau)}_{3e_{-}}$. In each case, this effectively doubles the separation between the Fock components while keeping the MPJC  interaction order fixed.

In general, the superposition coefficients produced by the MPJC protocol need not coincide exactly with those of the target binomial codeword. In such cases, the intermediate oscillator state can, in principle, be deterministically mapped onto the desired code state by applying a suitable bosonic unitary acting within the relevant finite-dimensional Fock subspace. These unitaries include number-selective phase operations (SNAP gates) and their generalizations, which provide full controllability over amplitudes and phases within a chosen Fock manifold and have been extensively demonstrated in circuit- and cavity-QED platforms~\cite{Heeres_PRL_2015}. We therefore treat this final single-mode operation as a controllable post-processing step and focus below on a representative example illustrating the procedure.

Consider the binomial state $\ket{\bar{0}}_{2,2} = (\ket{0} + \ket{4})/\sqrt{2}$. In the one-step MPJC protocol, generating this state requires a multiphoton order $m = 4$. Using the two-step protocol, however, the same state can be realized with a reduced multiphoton order $m = 2$. To obtain this target from $\ket{\psi_o(\tau)}_{3g_{+}}$, it suffices to set $n_1 = 0$; for $\ket{\psi_o(\tau)}_{3g_{-}}$ or $\ket{\psi_o(\tau)}_{3e_{+}}$, one should set $n_1 = 2$; and for $\ket{\psi_o(\tau)}_{3e_{-}}$, $n_1 = 4$. Substituting the corresponding $n_1$ and $m$ values into Eqs.~\eqref{eqn_psit_3g_both} and \eqref{eqn_psit_3e_both} produces oscillator states that are superpositions of $\ket{0}$, $\ket{2}$, and $\ket{4}$. The coefficient of the intermediate Fock state $\ket{2}$ in each case is given by
\begin{subequations}
\begin{align}
    C_{1g}^{(+)} &= \cos\theta\sin\phi\,\cos\tau_g - \sin\theta\cos\phi\,\sin\tau_g,    \\
    C_{1g}^{(-)} &= \cos\theta\cos\phi\,\cos\tau_g - \sin\theta\sin\phi\,\sin\tau_g,    \\
    C_{1e}^{(+)} &= \sin\theta\cos\phi\,\cos\tau_e + \cos\theta\sin\phi\,\sin\tau_e,    \\
    C_{1e}^{(-)} &= \cos\theta\cos\phi\,\sin\tau_e + \sin\theta\sin\phi\,\cos\tau_e, 
\end{align}
\end{subequations}
where $\tau_g = \sqrt{2}\tau$ and $\tau_e=\sqrt{12}\tau$.
The next step is to numerically optimize the control parameters ($\theta, \phi, \tau$) so that these coefficients vanish up to a desired numerical precision.

\begin{figure*}
    \centering
    \includegraphics[width=1.0\textwidth]{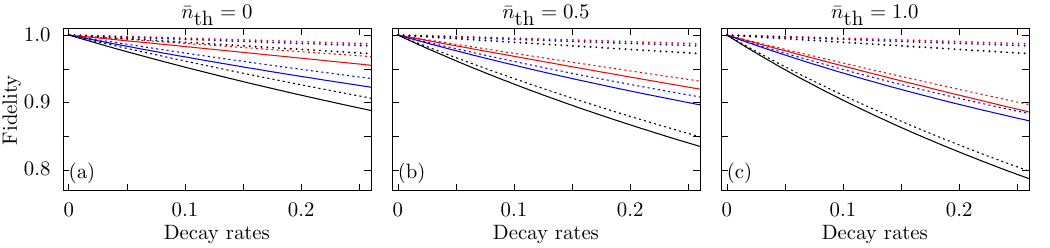}
    \vspace{-4ex}
    \caption{Fidelity degradation of the logical state $\ket{\bar{0}}_{2,2}$ as a function of the decay rates, under the influence of system–environment interactions. Panels (a)–(c) correspond to increasing thermal bath occupations $\bar{n}_{\text{th}}$, as indicated above each panel. The values of $\theta$ and $\tau$ are fixed to those for which the fidelity first reaches unity (within a numerical precision of $10^{-6}$) in the ideal, dissipation-free case. Three dissipation scenarios are considered: (i) oscillator-only coupling (red), (ii) qubit-only coupling (blue), and (iii) joint coupling of both subsystems to a common bath (black). In each case, dashed and dotted lines denote the effects of pure dissipation and pure dephasing, respectively, while solid lines represent the combined influence of both channels.}
    \label{fig:decoh}
\end{figure*}

The resulting optimal parameter combinations, for which $C_{1g}^{(\pm)}$ and $C_{1e}^{(\pm)}$ vanish within a tolerance of $10^{-4}$, are shown as scatter plots in Fig.~\ref{fig:c2c3}. A large number of valid solutions highlight the flexibility of the protocol. Selecting any such combination yields oscillator states of the form $\alpha \ket{0} + \beta \ket{4}$. If one of the resulting amplitude combinations $(\alpha, \beta)$ does not match the target values, a final unitary operation needs to be applied to transform the intermediate state into the desired binomial state with high fidelity.

Before proceeding, we emphasize that the scheme described in this section applies exclusively to code states comprising superpositions of two Fock states. Moreover, the reduction in the required $m$ value by a factor of two is a one-time improvement; further reductions by factors of 4, 8, and so on are not supported within this approach.

\section{Environmental effects}
\label{sec_env}

Until now, our analysis has focused exclusively on ideal unitary evolution of the bipartite quantum system, neglecting the critical effects of system–environment interactions.
In this section, we extend the study to include these effects by numerically simulating the full open-system state-preparation dynamics. Specifically, we replace the unitary MPJC evolution with a Lindblad master equation that accounts for both qubit relaxation and oscillator photon loss, such that the codeword is generated during the dissipative evolution rather than being prepared unitarily and subsequently propagated under dissipation
%In the following, we extend the study to incorporate these effects through numerical simulations, examining how environmental couplings affect the fidelity of the code states.
We adopt the standard Lindblad formalism, which relies on the Markovian approximation and other common assumptions such as the Born and secular approximations. Within this framework, the evolution of the reduced density matrix of the bipartite system, $\rho_{S}(t)$, is governed by the Lindblad master equation
\begin{align}
 \displaystyle\frac{d\rho_{S}}{dt}  = -i \left[H, \rho_{S}\right] + \sum_{k} \left( L_{k} \rho_{S} L_{k}^{\dagger} -  \frac{1}{2}\left[\rho_{S}, L_{k}^{\dagger} L_{k}\right] \right).
 \label{rhot_decoh}
\end{align}
Here, $L_k = \sqrt{\lambda_k}\, A_k$ are the collapse (Lindblad) operators, where the environment couples to the system through operators $A_k$ with corresponding rates $\lambda_k$. To simplify the numerical analysis, we assume that both the oscillator and the two-level system interact with a common thermal environment at a temperature characterized by the mean thermal occupation $\bar{n}_{\text{th}}$.

For the bosonic mode, the relevant Lindblad operators are $\sqrt{\lambda_{os_r}(1+\bar{n}_{\text{th}})}\, a$, $\sqrt{\lambda_{os_a} \bar{n}_{\text{th}}}\, a^{\dagger}$, and $\sqrt{\lambda_{os_d}}\, a^\dagger a$ which lead to emission/relaxation, absorption, and dephasing effects, respectively. Similarly, for the two-level system, the corresponding Lindblad operators are $\sqrt{\lambda_{qb_r}(1+\bar{n}_{\text{th}})}\, \sigma_-$, $\sqrt{\lambda_{qb_a} \bar{n}_{\text{th}}}\, \sigma_+$, and $\sqrt{\lambda_{qb_d}}\, \sigma_z$. While both relaxation and absorption rates depend on the bath temperature $\bar{n}_{\text{th}}$, the dephasing rates are temperature-independent. For simplicity, we assume $\lambda_{os_r}=\lambda_{os_a}$, and $\lambda_{qb_r}=\lambda_{qb_a}$.

The control parameters ($\theta$, $\phi$, $\tau$) used in the open-system simulations are taken from the unitary optimization discussed in Secs.~III–VI. Although these parameters are not re-optimized in the presence of dissipation, they provide a physically meaningful baseline for assessing how environmental decoherence degrades the performance of the protocol relative to the ideal unitary case. 
%A full re-optimization in the open-system setting is possible but lies beyond the scope of the present work.

To illustrate the impact of system–environment interactions on state preparation, we consider a representative example: the logical codeword $\ket{\bar{0}}_{2,2}$. While our analysis focuses on this specific case, we have verified that qualitatively similar behavior is observed for other logical states constructed as superpositions of two Fock states. To streamline the analysis and mitigate the effects of decoherence, we fix the parameter $\theta$ to the value at which the fidelity first reaches unity (within a numerical threshold of $10^{-6}$) in the absence of dissipation. This choice minimizes additional infidelity arising from the interaction with the environment. For the state $\ket{\bar{0}}_{2,2}$, we find numerically that this occurs at $\tau = 0.037699$ with the corresponding $\theta = 1.432566$.

\begin{figure*}
    \includegraphics[width=1.0\textwidth]{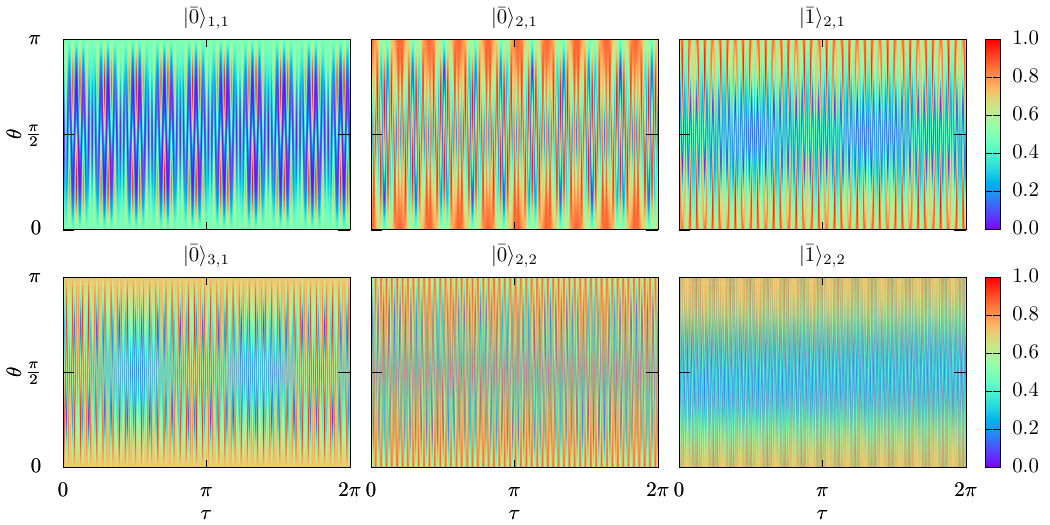}
    \caption{Fidelity between the time-evolved oscillator state $\rho_o(\tau)$ [as given in Eq.~\eqref{eqn_rho_t_det}] and the corresponding target binomial codewords $\ket{\bar{0}}{1,1}$, $\ket{\bar{0}}_{2,1}$, $\ket{\bar{1}}_{2,1}$, $\ket{\bar{1}}_{3,1}$, $\ket{\bar{0}}_{2,2}$, and $\ket{\bar{1}}{2,2}$, plotted as a function of the evolution time $\tau$ and the qubit superposition angle $\theta$. In contrast to the probabilistic protocol, no pair of $(\tau, \theta)$ achieves unit fidelity within a numerical tolerance of $10^{-4}$, highlighting the constrained parameter space inherent in the deterministic generation scheme (see main text). }
    \label{fig_two_sup_deter}
\end{figure*}

We first consider the case of zero temperature ($\bar{n}_{\text{th}} = 0$), where decoherence arises exclusively from vacuum fluctuations—i.e., energy relaxation and dephasing in the absence of thermal excitations [see Fig.~\ref{fig:decoh}(a)]. The relevant collapse operators governing the open-system dynamics are $\sqrt{\lambda_{\text{os}_r}}\, a$ and $\sqrt{\lambda_{\text{os}_d}}\, a^\dagger a$ for the oscillator, and $\sqrt{\lambda_{\text{qb}_r}}\, \sigma-$ and $\sqrt{\lambda_{\text{qb}_d}}\, \sigma_z$ for the qubit. We analyze three distinct scenarios: (i) the oscillator is coupled to the environment (red curves), (ii) the qubit is coupled (blue curves), and (iii) both subsystems are coupled simultaneously (black curves). In all cases, dashed and dotted lines correspond to pure dissipation and pure dephasing, respectively, while solid lines represent the combined influence of both. Notably, dephasing alone leads to a much lesser fidelity loss than energy relaxation for equivalent rates. Furthermore, the fidelity is more severely degraded when the qubit is coupled to the bath, as compared to the case in which the oscillator only interacts with the environment.

The effect of finite thermal excitations ($\bar{n}_{\text{th}} > 0$) is illustrated in Figs.~\ref{fig:decoh}(b) and \ref{fig:decoh}(c), where all relevant Lindblad operators actively contribute to the system’s open dynamics. As expected, the pure dephasing curves remain identical across panels, reflecting their insensitivity to thermal excitations. In contrast, increasing the thermal occupation number $\bar{n}_{\text{th}}$ leads to a progressive reduction in fidelity for all remaining cases, underscoring the detrimental role of thermal noise on the robustness of the encoded state.

For code states involving superpositions of a larger number of Fock states, the detrimental impact of losses is expected to be more pronounced. This is due to the fact that generating such states requires multiple sequential implementations of the MPJC interaction, with each stage introducing additional exposure to decoherence. As a result, losses compound over the course of the protocol, leading to a cumulative degradation in fidelity. However, by carefully selecting an optimal interaction time at each step--preferably at the earliest instance when high fidelity is achieved-- the effects of losses can be mitigated to a certain extent.

%%%%%%%%%%%%%%%%%%%%%%%%%%%%%%%%%%%%%%%%%%%%%%%%%%%%%%%%%%%%%%%%%%%%%%%%%%%%%%%%%%%%%%%%%%%%%%%%%%%%%%%%%%%%%%

\section{\label{sec_deterministic}Deterministic generation of binomial code states}
In this section, we present the deterministic state-generation protocol, focusing on the two- and three-Fock-component states. In this protocol, instead of performing a projective measurement on the qubit, the oscillator state is obtained by tracing out the qubit subsystem from the joint system state. This operation inevitably reduces the purity of the oscillator state due to the loss of coherence arising from qubit-oscillator entanglement.

For the state $\ket{\Psi(\tau)}$ in Eq.\eqref{eqn_psit_2fock}, the reduced density matrix of the oscillator, expressed in the truncated Fock basis ${\ket{n_1 - m}, \ket{n_1}, \ket{n_1 + m}}$, takes the form
\begin{align}
 \rho_{o}(\tau) = \begin{pmatrix}
        \vert x_4\vert^2 & x_2 x_4^* & x_1 x_3^* \\
        x_2^* x_4 & \vert x_1\vert^2 + \vert x_2\vert^2 & 0  \\
        x_1^* x_3 & 0 & \vert x_3\vert^2 
    \end{pmatrix}.
    \label{eqn_rho_t_det}
\end{align}

\begin{table}[ht]
\caption{\label{tab:table1}%
Optimal values of the parameters $\theta_{\mathrm{opt}}$ and $\tau_{\mathrm{opt}}$ at which each corresponding binomial code state achieves maximum fidelity in the deterministic generation protocol.}
\begin{ruledtabular}
\begin{tabular}{ccccc}
State &  $\theta_{\text{opt}}$ & $\tau_{\text{opt}}$ & Max. Fidelity \\
\colrule
\colrule
$\ket{\bar{0}}_{2,2}$  & 2.356194 &  3.524866 & 0.998582 \\
$\ket{\bar{0}}_{2,3}$  & 1.049291 & 5.441238  & 0.999613 \\
$\ket{\bar{1}}_{2,3}$  & 2.620088 & 3.229557  & 0.999473 \\
$\ket{\bar{0}}_{2,4}$  & 2.356194 & 3.229557  & 0.999052 \\
$\ket{\bar{0}}_{3,3}$  & 2.092300 & 4.856902  & 0.999700 \\
$\ket{\bar{1}}_{3,3}$  & 0.521504 & 0.772831  & 0.999763 \\
\end{tabular}
\end{ruledtabular}
\end{table}

\begin{table}[ht]
\caption{\label{tab:table2}%
Optimal values of the parameters $\theta_{\mathrm{opt}}$,$\phi_{\mathrm{opt}}$, and $\tau_{\mathrm{opt}}$ at which each corresponding binomial code state achieves maximum fidelity in the deterministic generation protocol.}
\begin{ruledtabular}
\begin{tabular}{ccccc}
State &  $\theta_{\text{opt}}$  &  $\phi_{\text{opt}}$ & $\tau_{\text{opt}}$ & Max. Fidelity \\
\colrule
\colrule
$\ket{\bar{0}}_{2,4}$  & 0.314159 & 1.897522 & 4.787787 & 0.991509 \\
$\ket{\bar{0}}_{2,5}$  & 0.502655 & 1.771858  & 6.170088 & 0.978205 \\
$\ket{\bar{1}}_{2,5}$  & 2.965663 & 0.967611  & 5.554336 & 0.987996 \\
$\ket{\bar{1}}_{2,6}$  & 2.827433 & 1.244071  & 5.554336 & 0.950207 \\
\end{tabular}
\end{ruledtabular}
\end{table}

A particularly favorable condition for deterministic generation occurs at specific combinations of the evolution time $\tau$ and the qubit superposition angle $\theta$ for which the amplitudes $|x_1| \approx 0$ and $|x_3| \approx 0$. Under these conditions, the reduced oscillator state approximates a nearly pure state of the form
\begin{align}
\ket{\psi_o(\tau)} \approx x_4(\tau) \ket{n_1 - m} + x_2(\tau) \ket{n_1},
\end{align}
which can be directly mapped (up to a local unitary transformation on the oscillator) to a binomial codeword comprising a superposition of two Fock states by selecting appropriate values of $n_1$ and $m$, analogous to the probabilistic protocol. The task then reduces to identifying specific pairs of $\tau$ and $\theta$  that, under the constraint $|x_1(\tau)| \approx 0$ and $|x_3(\tau)| \approx 0$, yield the desired superposition with high fidelity.

This constraint fundamentally distinguishes the deterministic protocol from its probabilistic counterpart. In the probabilistic approach, $\tau$ and $\theta$ are unconstrained and can be freely optimized to maximize the fidelity, allowing for a broader parameter search space. In contrast, the deterministic method limits the viable parameter sets, which inevitably reduces the achievable fidelity due to this restricted optimization landscape (see Fig.~\ref{fig_two_sup_deter}). This trade-off is clearly manifested in our numerical analysis, where no combination of $(\tau,\, \theta)$ was found to yield unit fidelity (within a numerical threshold of $10^{-4}$) for any of the six representative code states. The highest fidelities achieved in each case, along with the corresponding optimal parameter values, are summarized in Table~\ref{tab:table1}. It is important to note that these results pertain to the idealized, closed-system implementation of the deterministic protocol and do not include the effects of system–environment interactions. As expected, numerical simulations confirm that the inclusion of such interactions leads to a further degradation in fidelity.

Continuing to the case of three-component Fock superpositions, we proceed by setting $n_2=n_1+m$ in Eq.~\eqref{eqn_psit_3fock}, which yields the following bipartite state
\begin{align}
    \ket{\Psi(\tau)} &= x_1 \ket{g, n_1} + (x_2+y_4) \ket{e, n_1} \nonumber \\
    &\,\,\, + (x_3+y_1)\ket{g, n_1+m} +  y_2 \ket{e, n_1+m}   \nonumber\\
    &\,\,\, + x_4\ket{e, n_1-m} + y_3\ket{g, n_1+2m} . 
    \label{eqn_psit_3fock_ex}
\end{align}
After tracing out the qubit subsystem from $\ket{\Psi(\tau)}$ in Eq.\eqref{eqn_psit_3fock}, the reduced density matrix of the oscillator $\rho_{o}(\tau)$, can be expressed in the truncated Fock basis ${\ket{n_1 - m}, \ket{n_1}, \ket{n_1 + m}}, \ket{n_1 + 2m}$. It is given by
\begin{widetext}
    \begin{align}
    \rho_{o}(\tau) =
    \begin{pmatrix}
        \vert x_4\vert^2 & (x_2+y_4) x_4^* & x_4^* y_2 & 0 \\
        (x_2^*+y_4^*) x_4 & \vert x_1\vert^2 + \vert x_2+y_4\vert^2 & x_1^* (x_3+y_1) + (x_2^*+y_4^*) y_2 & x_1 y_3^* \\
        x_4^* y_2 & x_1 (x_3^*+y_1^*) + (x_2+y_4) y_2^*  & \vert x_3+y_1\vert^2 + \vert y_2\vert^2 & (x_3^*+y_1^*)y_3\\
        0 & x_1^*y_3 & (x_3^*+y_1^*)y_3 & \vert y_3\vert^2
    \end{pmatrix}.
    \label{eqn_rho_t_det_3sup}
\end{align} 
\end{widetext}
Now, in the limit, $x_1\approx 0$, $x_3\approx 0$, $y_1(\tau)\approx 0$, $y_3(\tau)\approx 0$, the above density matrix assumes the pure state of the form
\begin{align}
    \ket{\psi_o(\tau)} \approx x_4\ket{n_1 - m} + (x_2+y_4) \ket{n_1} + y_2\ket{n_1+m}.
\end{align}
As before, this state can be mapped (up to a local unitary transformation on the oscillator) to a binomial codeword comprising a superposition of three Fock states by appropriately choosing the values of $n_1$ and $m$. However, due to constraints on the coefficients, it is not possible to attain near-unity fidelity for the desired code states. Representative best-case scenarios for the four states considered in this study are summarized in Table~\ref{tab:table2}.

%%%%%%%%%%%%%%%%%%%%%%%%%%%%%%%%%%%%%%%%%%%%%%%%%%%%%%%%%%%%%%%%%%%%%%%%%%%%%%%%%%%%
\section{Concluding remarks}
\label{sec_conc}
%Standard JC is insufficient.
%\textcolor{red}{\bf Peter:''Last point, maybe I had mentioned this before: The binomial codes are rotation-symmetric codes. As such it may be possible to generate the states from "standard" JC interactions in the dispersive regime, also by multiple applications of such unitaries. We have to comment on this. Check Refs."}~\cite{Grimsmo_PRX_2020,Li_AdvQTech_2023,Li_AdvQTech_2024}

Rotation-symmetric bosonic codes (RSBCs) encode quantum information in states invariant under discrete phase-space rotations, leveraging rotation symmetries to enhance error resilience and fault tolerance~\cite{Grimsmo_PRX_2020,Li_AdvQTech_2023,Li_AdvQTech_2024}. Various families of RSBCs have been proposed, including (squeezed) cat codes~\cite{Liu_PRA_2005,Leghtas_PRL_2013,Mirrahimi_NJP_2014,Bergmann_PRA_2016,Li_PRL_2017,Schlegel_PRA_2022}, Pegg–Barnett codes~\cite{Baragiola_PRL_2019}, GKP-like codes~\cite{GKP_PRA_2001,Albert_PRA_2018}, and binomial codes~\cite{Albert_PRA_2018,Juliette_2024}. In this work, we focus on the binomial code family, which consists of finite superpositions of Fock states with coefficients drawn from a binomial distribution.

%xu_2022_arxiv

We have presented a general and high-fidelity scheme for synthesizing {\em arbitrary} binomial codewords by harnessing nonlinear multiphoton interactions between a bosonic mode (oscillator) and a two-level system (qubit). Our approach is based on the coherent dynamics of the multiphoton Jaynes–Cummings (MPJC) Hamiltonian and assumes experimentally reasonable capabilities: initialization of the oscillator and qubit, access to multiphoton interactions, and qubit measurements during the evolution. While iterative methods based on standard cavity QED interactions have proven effective for generating cat codes starting from easily prepared coherent states, a similar strategy is less practical for binomial codes. As discussed in Appendix~\ref{sce:RSBC}, the primitive states required for binomial codewords are themselves complex superpositions of Fock states and do not offer a clear experimental advantage over direct codeword generation.

A key insight from our analysis is the essential role of a coherent initial qubit state in generating superpositions of target Fock states. This requirement stems from the inability of the MPJC Hamiltonian to induce coherence in either subsystem if the joint initial state lacks it—a constraint highlighted in recent studies~\cite{laha_AdvQT_2023,laha_PRR_2024}. More broadly, the challenge of harnessing noisy or incoherent resources for generating entanglement and coherence remains an active area of investigation~\cite{laha_oe_2022,laha_josab_2023,laha_PRL_2022}. This raises an important question: can the inclusion of an additional dispersive spin-boson interaction compensate for the coherence shortfall of MPJC dynamics and enable high-fidelity synthesis of bosonic code states using noisy initial resources? The authors aim to address this issue in the future.

Another important feature in our scheme is the ability to reduce the required multiphoton transition order by a factor of two without sacrificing fidelity, thus significantly enhancing the experimental feasibility of the protocol. This one-time improvement is achieved by judiciously selecting the initial oscillator state and exploiting coherent interference effects. 

In developing our protocol, we explored two complementary approaches to generating binomial codewords. The probabilistic protocol, which relies on projective measurement of the qubit after joint evolution, offers higher fidelities and greater flexibility in scanning the parameter space, making it better suited for precise codeword synthesis. 
%However, a key drawback of this approach is its inherent probabilistic nature—only a subset of measurement outcomes yield the desired oscillator state, resulting in a trade-off between fidelity and generation efficiency.
In contrast, the deterministic protocol, which involves tracing out the qubit degrees of freedom without any measurement, is operationally simpler and potentially more suitable in scenarios where projective measurements are inefficient, technically challenging, or resource-intensive. Due to the constrained parameter space accessible in the deterministic setting, this protocol typically achieves slightly lower fidelities. 
%A detailed comparison, along with results for the deterministic protocol, is provided in Appendix~\ref{sec_deterministic}.

Our results contribute to the broader effort of realizing fault-tolerant bosonic quantum computing, not only by identifying an efficient method for preparing logical codewords, but also by highlighting the utility of nonlinear light–matter interactions in continuous-variable quantum systems. The methods outlined here are versatile and can be extended to prepare larger classes of bosonic codewords, potentially including cat codes and other non-Gaussian encodings.

In summary, this work provides a scalable and practical pathway toward high-fidelity generation of binomial code states, addressing a critical bottleneck in the realization of bosonic quantum error correction.
%.

\begin{acknowledgments}
We thank P. A. Ameen Yasir and S. Siddardha Chelluri for valuable discussions. We acknowledge funding from the BMBF/BMFTR in Germany (QR.X/QR.N, PhotonQ, QuKuK, QuaPhySI), from the Deutsche Forschungsgemeinschaft (DFG, German Research Foundation) – Project-ID 429529648 – TRR 306 QuCoLiMa (Quantum Cooperativity of Light and Matter), and also from the EU project CLUSTEC (Grant Agreement No. 101080173). 
\end{acknowledgments}

\appendix
\section{\label{sec_app_state_vector}Details on the state vector}
In this Appendix, we present additional details on the oscillator state vectors obtained through our protocol for generating binomial codewords involving superpositions of two and three Fock states. Furthermore, we extend our analysis to include a representative example of a superposition of four Fock states, illustrating the scalability of the protocol to more complex codewords.

\subsection{Superposition of two Fock states}
Applying the time-independent Schr\"odinger equation,
$i \frac{d}{dt}\ket{\Psi(t)} = H_{\text{int}} \ket{\Psi(t)}$, with $H_{\text{int}}$ and $\ket{\Psi(t)}$ as defined in Eqs.~\eqref{eqn_H_int} and \eqref{eqn_psit_2fock}, respectively,  we get four time-dependent coefficients that obey the following coupled differential equations
\begin{subequations}
\begin{align}
   \dot{x}_1(\tau) &= -\sqrt{n_1!/(n_1-m)!}\, x_4(\tau), \\
   \dot{x}_2(\tau) &= \sqrt{(n_1+m)!/n_1!}\, x_3(\tau), \\
   \dot{x}_3(\tau) &= -\sqrt{(n_1+m)!/n_1!}\, x_2(\tau), \\
   \dot{x}_4(\tau) &= \sqrt{n_1!/(n_1-m)!}\, x_1(\tau), 
\end{align}
\label{eqn_xi_dots}
\end{subequations}
for $n_1\geqslant m$. If $n_1< m$, $\dot{x}_1(\tau) =0$ and $x_4(\tau) =0$. Note that $\tau=gt$.
The initial condition in Eq.\eqref{eqn_psi0} implies $x_1(0) = \cos\theta$, $x_2(0) = \sin\theta$, and $x_3(0) = x_4(0) = 0$. Solving the coupled differential equations with these initial values yields explicit expressions for all four time-dependent coefficients. The resulting solutions are provided in Eq.~\eqref{eqn_xt_2fock} of the main text.

\subsection{Superposition of three Fock states}
Following the same procedure as outlined earlier, and using the state $\ket{\Psi(t)}$ defined in Eq.\eqref{eqn_psit_3fock}, we obtain a set of eight coupled differential equations describing the system's time evolution. The first four equations are identical to the ones presented in Eq.\eqref{eqn_xi_dots}, while the remaining four are given by
\begin{subequations}
\begin{align}
%   i \dot{x}_1 &= \sqrt{\frac{n_1!}{(n_1-m)!}} x_4, \quad   i \dot{x}_2 = \sqrt{\frac{(n_1+m)!}{n_1!}} x_3, \\
%   i \dot{x}_3 &= \sqrt{\frac{(n_1+m)!}{n_1!}} x_2, \quad   i \dot{x}_4 = \sqrt{\frac{n_1!}{(n_1-m)!}} x_1,  \\
   \dot{y}_1(\tau) &= -\sqrt{n_2!/(n_2-m)!}\, y_4(\tau), \\
   \dot{y}_2(\tau) &= \sqrt{(n_2+m)!/n_2!}\, y_3(\tau), \\
   \dot{y}_3(\tau) &= -\sqrt{(n_2+m)!/n_2!}\, y_2(\tau), \\
   \dot{y}_4(\tau) &= \sqrt{n_2!/(n_2-m)!}\, y_1(\tau).
\end{align}
\label{eqn_yi_dots}
\end{subequations}
The initial conditions are given by $x_1(0) = \cos\theta\cos\phi$, $x_2(0) = \sin\theta\cos\phi$, $y_1(0) = \cos\theta\sin\phi$, and $y_2(0) = \sin\theta\sin\phi$, with all other coefficients vanishing initially: $x_3(0) = x_4(0) = y_3(0) = y_4(0) = 0$. Solving the resulting set of coupled differential equations yields all the eight coefficients and are given in Eq.~\eqref{eqn_xyt_3fock_short}. 
\subsection{Superposition of four and higher Fock states}
In the main text, we showed that binomial codewords composed of superpositions of two or three Fock states can be engineered by choosing an appropriate initial state, selecting the multiphoton order $m$, and optimizing the control parameters ($\theta$, $\phi$, and $\tau$). Here, we extend this construction to superpositions involving larger numbers of Fock components by systematically applying the same procedure.

%\subsubsection{Superposition of four Fock states}
First, consider binomial states composed of four Fock components. A simple algebraic count shows  that there exist eleven logical $\ket{\bar{0}}$ states (five for $K=6$ and six for $K=7$) and thirteen logical $\ket{\bar{1}}$ states (six for $K=7$ and seven for $K=8$). The corresponding generic forms of these four-component logical states are
\begin{subequations}
\begin{align}
\ket{\bar{0}}_{N,6} &=\displaystyle \frac{1}{4\sqrt{2}} \big(\ket{0} + \sqrt{15} \ket{2N} + \sqrt{15} \ket{4N} + \ket{6N}\big), \\
\ket{\bar{0}}_{N,7} &=\displaystyle \frac{1}{8}\big(\ket{0} + \sqrt{21}\ket{2N} + \sqrt{35}\ket{4N} + \sqrt{7}\ket{6N}\big), \\
\ket{\bar{1}}_{N,7} &=\displaystyle \frac{1}{8}\big(\sqrt{7} \ket{N} + \sqrt{35}\ket{3N} + \sqrt{21}\,\ket{5N}+\ket{7N}\big), \\
\ket{\bar{1}}_{N,8} &=\displaystyle \frac{1}{4}\big(\ket{N} + \sqrt{7}\ket{3N} + \sqrt{7}\ket{5N} + \ket{7N}\big) .
\end{align}
\end{subequations}

Direct generation of these four-Fock-component target binomial states from Eqs.~\eqref{eqn_psi_t_g_3} (ground-state postselection)  and \eqref{eqn_psi_t_e_3} (excited-state postselection)  requires appropriate choices of $n_1$, $n_2$, and $m$, as summarized in Table~\ref{tab:four_fock_params}. Note that for each logical state, two parameter combinations are possible; from an experimental standpoint, it is preferable to choose the combination with the smaller $m$. Once $n_1$, $n_2$, and $m$ are fixed, the next step is to identify the control parameters $(\theta, \phi, \tau)$ that maximize the fidelity to the target state. We performed this numerical serach for two representative cases, $\ket{\bar{0}}_{N,6}$ and $\ket{\bar{1}}_{2,7}$. Remarkably, for both states—and for all admissible choices of $m$ and postselection—we found that only a small number of parameter triplets (between 4 and 26) achieve fidelity within a numerical tolerance of $10^{-4}$, and none met the stricter threshold of $10^{-6}$.

%, with the results presented in Fig.~\ref{}.

%For example, to obtain the target logical $\ket{\bar{0}}$ (similarly, $\ket{\bar{1}}$) states Eq.~\eqref{eqn_psi_t_g_3} , there exists two possibilities $\{n_1, n_2, m\} = \{0, 4N, 2N\}$ or $\{0, 2N, 4N\}$ (equivalently, $\{n_1, n_2, m\} = \{N, 5N, 2N\}$ or $\{N, 3N, 4N\}$).  On the other hand, to obtain the target logical $\ket{\bar{0}}$ (similarly, $\ket{\bar{1}}$) states Eq.~\eqref{eqn_psi_t_e_3} , there exists two possibilities $\{n_1, n_2, m\} = \{2N, 6N, 2N\}$ or $\{4N, 6N, 4N\}$ (equivalently, $\{n_1, n_2, m\} = \{3N, 7N, 2N\}$ or $\{5N, 7N, 4N\}$)
\begin{table}[ht]
\centering
\caption{Parameter sets $\{n_1, n_2, m\}$ for generating four-Fock-component logical states via Eqs.~\eqref{eqn_psi_t_g_3} (ground-state postselection) and \eqref{eqn_psi_t_e_3} (excited-state postselection).}
\label{tab:four_fock_params}
\begin{ruledtabular}
\begin{tabular}{c c c c c}
Postselection & Logical state & $n_1$ & $n_2$ & $m$  \\
\hline
\multirow{4}{*}{$\ket{g}$}  & \multirow{2}{*}{$\ket{\bar{0}}$} & 0 & 4N & 2N \\
										      & & 0 & 2N & 4N \\
\cline{2-5}
					& \multirow{2}{*}{$\ket{\bar{1}}$} & N & 5N & 2N \\ 
										       & & N & 3N & 4N \\
\hline
\multirow{4}{*}{$\ket{e}$ } & \multirow{2}{*}{$\ket{\bar{0}}$} & 2N & 6N & 2N \\
											& & 4N & 6N & 4N \\
\cline{2-5}
					& \multirow{2}{*}{$\ket{\bar{1}}$} & 3N & 7N & 2N \\
											& & 5N & 7N & 4N \\
\end{tabular}
\end{ruledtabular}
\end{table}

%To obtain target binomial states comprising superpositions of five or six Fock components, w

The limited number of viable parameter combinations for directly generating these four-component states can be mitigated by increasing the number of MPJC steps. To illustrate this, we begin by preparing the full system in the initial state
\begin{align}
    \ket{\Psi(0)} &= \left(\cos\theta\ket{g}+\sin\theta\ket{e}\right) \otimes\big(\sin\phi_1\cos\phi_2\ket{n_1}\nonumber\\
    &\quad+\sin\phi_1\sin\phi_2\ket{n_2}+\cos\phi_1\ket{n_3}\big).
\end{align}

The time-evolved state in this case vector takes the form
\begin{align}
\ket{\Psi(\tau)} &= \sum_{i=1}^{4} x_i(\tau)\,\ket{\chi_i(n_1)} + \sum_{i=1}^{4} y_i(\tau)\,\ket{\chi_i(n_2)},\nonumber\\
& + \sum_{i=1}^{4} z_i(\tau)\,\ket{\chi_i(n_3)}
\label{eqn_psit_4fock}
\end{align}
where $\ket{\chi_i(n)} \in {\ket{g,n},\ket{e,n},\ket{g,n+m},\ket{e,n-m}}$.

Proceeding as before, we now derive the set of twelve coupled differential equations governing the system dynamics. The initial conditions are given by
$x_1(0) = \cos\theta \sin\phi_1 \cos\phi_2$,
$x_2(0) = \sin\theta \sin\phi_1 \cos\phi_2$,
$y_1(0) = \cos\theta \sin\phi_1 \sin\phi_2$,
$y_2(0) = \sin\theta \sin\phi_1 \sin\phi_2$,
$z_1(0) = \cos\theta \cos\phi_1$,
$z_2(0) = \sin\theta \cos\phi_1$,
with all remaining coefficients initialized to zero:
$x_3(0) = x_4(0) = y_3(0) = y_4(0) = z_3(0) = z_4(0) = 0$.
The time-dependent amplitudes factorize as
\begin{subequations}
\begin{align}
x_i(\tau) &= \sin\phi_1 \cos\phi_2\, x_i^{(2)}(n_1,m,\theta, \tau),\\
y_i(\tau) &= \sin\phi_1 \sin\phi_2 \, x_i^{(2)}(n_2,m,\theta, \tau),\\
z_i(\tau) &= \cos\phi_1 \, x_i^{(2)}(n_3,m,\theta, \tau),
\end{align}
\label{eqn_xyt_4fock_short}
\end{subequations}
where $x_i^{(2)}$ denotes the coefficients of the two-Fock solution given previously in Eq.~\eqref{eqn_xt_2fock}. Thus the temporal evolution decomposes naturally into three independent excitation manifolds, weighted by $\sin\phi_1\cos\phi_2$, $\sin\phi_1\sin\phi_2$, and $\cos\phi_1$, respectively. For $n_1<m$, $x_1(\tau) = \cos\theta \sin\phi_1 \cos\phi_2, \quad  x_4(\tau) = 0$, for $n_2<m$, $y_1(\tau) = \cos\theta\sin\phi_1\sin\phi_2, \quad  y_4(\tau) = 0$, and  for $n_3<m$, $z_1(\tau) = \cos\theta\cos\phi_1, \quad  z_4(\tau) = 0$. 
%The two-Fock solution is recovered simply by setting $\phi_1=\phi_2=0$.

Next, subject to a projective qubit ground and excited state measurements, the (unnormalized) oscillator states reduces to
\begin{subequations}
    \label{eqn_psi_t_4}
    \begin{align}
    \label{eqn_psi_t_g_4}
    \ket{\psi_o(\tau)}_g &= x_1 \ket{n_1} + x_3 \ket{n_1+m}  + y_1 \ket{n_2} + y_3 \ket{n_2+m} \nonumber\\
   				 &\qquad + z_1 \ket{n_3} +  z_3 \ket{n_3+m},\\
    \ket{\psi_o(\tau)}_e &= x_2 \ket{n_1} + x_4 \ket{n_1-m}  + y_2 \ket{n_2} + y_4 \ket{n_2-m} \nonumber\\
   				 &\qquad + z_2 \ket{n_3} +  z_4 \ket{n_3-m}.
    \label{eqn_psi_t_e_4}
\end{align}
\end{subequations}
% \begin{align}
%     \ket{\psi_o(\tau)}_g &= x_1(\tau) \ket{n_1} + x_3(\tau) \ket{n_1+m}  + y_1(\tau) \ket{n_2}\nonumber\\
%    				 & + y_3(\tau) \ket{n_2+m} + z_1(\tau) \ket{n_3} +   z_3(\tau) \ket{n_3+m}.
%     \label{eqn_psi_t_g_4}
% \end{align}

There are multiple ways in which the six–Fock-component superposition in Eq.~\eqref{eqn_psi_t_4} can be reduced to four distinct components through appropriate choices of  $n_1$, $n_2$, and $m$. As an illustrative example, consider the logical state  $\ket{\bar{0}}_{2,6}$. One possible realization from Eq.~\eqref{eqn_psi_t_e_4} is obtained by choosing $n_1 = m = 4$, $n_2 = 8$, and $n_3 = 12$, which yields the following explicit form of the oscillator 
state
\begin{align}
 \ket{\phi} &=  \mathcal{N}\left( C_0 \ket{0} + C_1 \ket{4} + C_2 \ket{8} + C_{3} \ket{12}\right), 
\end{align}
where
\begin{align}
 C_0 &= \cos\theta \sin{\phi_1} \cos{\phi_2} \sin\left( \sqrt{4!} \, \tau\right), \nonumber\\
 C_1 &= \sin\theta \sin{\phi_1} \cos{\phi_2} \cos\left(\sqrt{8!/4!}\, \tau\right) \nonumber\\
 &\quad+ \cos\theta \sin{\phi_1} \sin{\phi_2} \sin \left(\sqrt{8!/4!}\, \tau\right),\\
 C_2 &= \sin\theta \sin{\phi_1}S_{\phi_2} \cos\left(\sqrt{12!/8!}\, \tau\right) \nonumber\\
 &\quad+ \cos\theta \cos{\phi_1}  \sin\left(\sqrt{12!/8!}\, \tau\right), \\
 C_{3} &= \sin\theta \cos{\phi_1}\cos\left(\sqrt{16!/12!}\, \tau\right)\Big\},
\end{align}
and $\mathcal{N}$ the normalization factor.
% is given by
% \begin{equation}
%    \mathcal{N} = \left[C_0^2 +  C_4^2 + C_8^2 +  C_{12}^2 \right]^{-\frac{1}{2}}.
% \end{equation}
The fidelity between the state $\ket{\phi}$ and the target binomial codeword  $\ket{\bar{0}}_{2,6}$ is given by
\begin{align}
 F &= \tfrac{\mathcal{N}^2}{32}\left| C_0 + \sqrt{15}\, C_1 + \sqrt{15}\, C_2 + C_{3} \right|^2. 
\end{align}

As in previous cases, achieving unit fidelity requires numerical optimization over the parameter $\theta$, $\phi_1$, $\phi_2$, and the evolution time $\tau$. Specifically, we scanned $\theta$, $\phi_1$, and $\phi_2$ over the interval $[0, \pi]$ using 251 uniformly spaced points for each, and $\tau$ over $[0, 2\pi]$ with 501 points. Within this parameter space, we identified approximately 5024 parameter combinations for which the fidelity approaches unity to within a numerical precision of $10^{-4}$, and six combinations that satisfy a more stringent threshold of $10^{-6}$. This demonstrates a clear increase in the number of viable parameter choices, albeit at the cost of an additional MPJC step.

%\subsubsection{Superposition of five and six Fock states}
On the other hand, the six-Fock–component superposition state in Eq.~\eqref{eqn_psi_t_g_4} can be reduced to five distinct components in exactly six different ways, namely by choosing any of the constraints $n_i = n_j \pm m$ where $i\ne j$ and $1\leqslant i,j\leqslant 3$.

%\paragraph*{Case (i): $n_2=n_1+m$.} 
In the case $n_2 = n_1 + m$, the accessible Fock ladder is
${k_0, k_1, k_2, k_3, k_4} = {n_1, n_1+m, n_1+2m, n_3,  n_3+m}$. For the alternative choice $n_2 = n_1 - m$ with $n_1 \ge m$, the ladder becomes
${n_1-m, n_1, n_1+m, n_3, n_3+m}$.
In both cases, the oscillator state derived from Eq.~\eqref{eqn_psi_t_g_4} can be written in the unified form
\begin{align}
	\label{eqn_psit_5g_both}
	\ket{\psi_o(\tau)}_{5g_{\pm}} = \mathcal{N}_{5g_\pm} \sum_{l=0}^4 C_{lg}^{(\pm)}(\tau, \theta, \phi_1, \phi_2)\, \ket{k_l}.
\end{align}
where $\mathcal{N}_{5g_\pm}$ is the normalization factor, and the exact expressions of the coefficients $C_{lg}^{(\pm)}$ can be straightforwardly obtained. 
%The normalizations $\mathcal{N}_{g_\pm}  = \left\{|C_{0g}^{(\pm)}|^2+|C_{1g}^{(\pm)}|^2+|C_{2g}^{(\pm)}|^2\right\}^{-1/2}$. 
Writing the target states in the form $\sum_{l}c_l\ket{k_l}$ for both cases, the fidelity for both constructions can be succinctly expressed in the form
\begin{equation}
	F_{5g_{\pm}}(\tau, \theta,\phi) = \frac{\left| \sum\limits_{l=0}^4 c_l C_{lg}^{(\pm)}(\tau, \theta, \phi_1, \phi_2)\right|^2}{ \sum\limits_{l=0}^4 \left|C_{lg}^{(\pm)}(\tau, \theta, \phi_1, \phi_2) \right|^2  }.
	\label{eqn_fid_three_fock_ket_5g_pm}
\end{equation}
The next step is to scan over the control parameters  $(\theta,\,\phi_1,\,\phi_2,\,\tau)$ and identify combinations for which the fidelity reaches unity within the desired numerical precision. We also note that the state in  Eq.~\eqref{eqn_psit_4fock} can, in principle, be used directly to generate binomial code states comprising six Fock components. Although we do not present an explicit example here, such a construction is straightforward in principle; the only limitation is that the corresponding 
optimization becomes increasingly demanding as the number of control parameters grows.

\section{\label{sec_two_fock_span}Controllability of two-Fock-state superpositions}
In this Appendix, we show that the states 
%$\ket{\psi_o(\tau)}_{2g} = \mathcal{N}_{2g}\!\left[\cos\theta\,\ket{n_1} - \sin\theta\,\sin(a_{n_1,m}\tau)\,\ket{n_1+m} \right]$ and $\ket{\psi_o(\tau)}_{2e} = \mathcal{N}_{2e}\!\left[\sin\theta\,\cos(a_{n_1,m}\tau)\,\ket{n_1}\right. \left. + \cos\theta\,\sin(b_{n_1,m}\tau)\,\ket{n_1-m} \right]$
\begin{align}
\ket{\psi_o(\tau)}_{2g} &= \mathcal{N}_{2g}\!\left[
\cos\theta\,\ket{n_1} - \sin\theta\,\sin(a_{n_1,m}\tau)\,\ket{n_1+m} \right], \nonumber\\[2pt]
\ket{\psi_o(\tau)}_{2e} &= \mathcal{N}_{2e}\!\left[\sin\theta\,\cos(a_{n_1,m}\tau)\,\ket{n_1} \right. \nonumber \\
&\qquad\qquad \left.+ \cos\theta\,\sin(b_{n_1,m}\tau)\,\ket{n_1-m} \right], \nonumber
\end{align}
enable arbitrary superpositions of the respective Fock-state pairs $\{\ket{n_1},\ket{n_1+m}\}$ and $\{\ket{n_1},\ket{n_1-m}\}$. In each case, the engineered interaction couples only a two-dimensional subspace of the oscillator Hilbert space, resulting in Rabi-type dynamics with number-dependent couplings $a_{n_1,m}=\sqrt{(n_1+m)!/n_1!}$ and $b_{n_1,m}=\sqrt{n_1!/(n_1-m)!}$.

For $\ket{\psi_o(\tau)}_{2g}$, the amplitude ratio is given  by
\begin{equation}
\frac{C_{n_1+m}}{C_{n_1}} = -\tan\theta\,\sin(a_{n_1,m}\tau).
\end{equation}
The above ratio can take any real value since $\tan\theta$ spans $(-\infty,\infty)$ and $\sin(a_{n_1,m}\tau)\in[-1,1]$, meaning that any real superposition of the form $C_1\ket{n_1} + C_2\ket{n_1+m}$ is directly reachable. If one additionally wishes to control the relative phase between the two components, this can be achieved by adjusting the phase of the external drive, which provides independent complex phase control of the superposition.

Similarly, for  $\ket{\psi_o(\tau)}_{2e}$, the ratio
\begin{equation}
\frac{C_{n_1-m}}{C_{n_1}} = \cot\theta\,\tan(b_{n_1,m}\tau),
\end{equation}
again covering all real values. This guarantees access to all normalized states $C_1\ket{n_1} + C_2\ket{n_1-m}$.

Therefore, by tuning $\theta$ and $\tau$, both manifolds provide full Bloch-sphere controllability within their respective two-Fock-state subspaces, enabling the preparation of arbitrary superpositions.

We mention in passing that although the states
$|\psi_o(\tau)\rangle_{2g}$ and $|\psi_o(\tau)\rangle_{2e}$ allow full controllability within their two-dimensional Fock subspaces, the specific parameters $(\theta,\tau)$ for a target superposition typically require a numerical search. This is because the effective couplings $a_{n_1,m}$ and $b_{n_1,m}$
% \[
% a_{n_1,m} = \sqrt{\frac{(n_1+m)!}{n_1!}}, \qquad
% b_{n_1,m} = \sqrt{\frac{n_1!}{(n_1-m)!}}
% \]
are generally irrational, preventing a simple analytical solution. Numerical optimization over $\theta$ and $\tau$ efficiently identifies the desired superposition with high fidelity.

%%%%%%%%%%%%%%%%%%%%%%%%%%%%%%%%%%%%%%%%%%%%%%%%%%%%%%%%%%%%%%%%%%%%%%%%%%%%%%%%%%%%%%
\section{\label{sce:RSBC}Binomial codewords via rotation of primitive states}
Following the framework developed in Ref.~\cite{Grimsmo_PRX_2020}, in this Appendix, we demonstrate explicitly how the logical binomial codewords defined in Eq.~\eqref{eqn_bonom_code_states} can be obtained from a single {\em primitive state} using parity-selective rotations. This construction places the binomial code within the broader class of RSBCs and highlights the underlying discrete rotational symmetries that structure these codes.

% \begin{align}
%     \ket{\bar{\mu}}_{N,S} = \frac{1}{\sqrt{2^N}}\sum_{k=0}^{\left\lceil\frac{{N+1}}{2}\right\rceil-\mu}\sqrt{\binom{N+1}{2k+\mu}} \ket{(S+1)(2k+\mu)},
%     \label{eqn_bonom_code_states_app}
% \end{align}

\subsection{Construction of binomial codewords from a primitive state}

We define the primitive state as a weighted superposition over Fock states spaced by $(S+1)$, with weights drawn from binomial coefficients
\begin{equation}
\ket{\Theta}_{N,K} = \frac{1}{\sqrt{2^{K-1}}} \sum_{k=0}^{K} \sqrt{\binom{K}{k}} \ket{Nk}.
\end{equation}
% \begin{equation}
% \ket{\Theta}_{N,S} = \frac{1}{\sqrt{2^N}} \sum_{m=0}^{N+1} \sqrt{\binom{N+1}{m}} \ket{(S+1)m}.
% \end{equation}

This state includes all Fock components required to build both logical codewords $\ket{\bar{0}}_{N,K}$ and $\ket{\bar{1}}_{N,K}$, since it contains both even and odd values of $k$.

To extract the individual logical codewords, we exploit the discrete rotation symmetry in Fock space using the operator
\begin{equation}
\hat{R}(\phi) = e^{i\phi \hat{n}}, \quad \hat{n} = \hat{a}^\dagger \hat{a}.
\end{equation}
In particular, a rotation by $\phi = \pi$ maps each Fock state as $\hat{R}(\pi)\ket{n} = (-1)^n \ket{n}$,
% \begin{equation}
% \hat{R}(\pi)\ket{n} = (-1)^n \ket{n},
% \end{equation}
thus separating the primitive state into parity components.

Define the logical codewords as
\begin{align}
\ket{\bar{0}}_{N,K} &= \mathcal{N}_0\left( \ket{\Theta}_{N,K} + \hat{R}(\pi) \ket{\Theta}_{N,K} \right), \\
\ket{\bar{1}}_{N,K} &= \mathcal{N}_1 \left( \ket{\Theta}_{N,K} - \hat{R}(\pi) \ket{\Theta}_{N,K} \right),
\end{align}
% \begin{align}
% \ket{\bar{0}}_{N,S} &= \mathcal{N}_0\left( \ket{\Theta}_{N,S} + \hat{R}(\pi) \ket{\Theta}_{N,S} \right), \\
% \ket{\bar{1}}_{N,S} &= \mathcal{N}_1 \left( \ket{\Theta}_{N,S} - \hat{R}(\pi) \ket{\Theta}_{N,S} \right),
% \end{align}
where $\mathcal{N}_{\mu}$ are normalization constants ($\mu = 0, 1$). These operations project onto even and odd values of $k$, respectively, thereby isolating the desired logical basis states.

This construction guarantees orthogonality of the logical states and yields the exact codeword forms prescribed by the binomial code definition.

\subsection{Example: binomial code for $N=3, S=3$}

As an illustrative example, consider the case where $N=3$ and $K=3$, leading to Fock spacings of $3$. As defined in Eq.~\eqref{eqn_sup_two_fock_states}, the binomial codewords are: $\ket{\bar{0}}_{3,3} = \frac{1}{2} \left( \ket{0} + \sqrt{3} \ket{6} \right)$, $\ket{\bar{1}}_{3,3} = \frac{1}{2} \left( \sqrt{3} \ket{3} + \ket{9} \right)$.
% These match the general formula:
% \begin{equation}
% \ket{\bar{\mu}}_{N,S} = \frac{1}{\sqrt{2^N}} \sum_{k=0}^{\left\lceil \frac{N+1}{2} \right\rceil - \mu} \sqrt{\binom{3}{2k + \mu}} \ket{3(2k + \mu)}.
% \end{equation}
The associated primitive state is:
\begin{equation}
\ket{\Theta}_{3,3} = \frac{1}{2} \left( \ket{0} + \sqrt{3} \ket{3} + \sqrt{3} \ket{6} + \ket{9} \right).
\end{equation}

Applying the parity-selective rotation $\hat{R}(\pi)$, we obtain
\begin{equation}
\hat{R}(\pi)\ket{\Theta}_{3,3} = \frac{1}{2} \left( \ket{0} - \sqrt{3} \ket{3} + \sqrt{3} \ket{6} - \ket{9} \right).
\end{equation}

Therefore, it is straightforward to extract the logical states as shown below
\begin{align}
\ket{\bar{0}}_{3,3} &= \frac{1}{\sqrt{4}} \left( \ket{\Theta}_{3,3} + \hat{R}(\pi) \ket{\Theta}_{3,3} \right) = \frac{1}{2} \left( \ket{0} + \sqrt{3} \ket{6} \right), \\
\ket{\bar{1}}_{3,3} &= \frac{1}{\sqrt{4}} \left( \ket{\Theta}_{3,3} - \hat{R}(\pi) \ket{\Theta}_{3,3} \right) = \frac{1}{2} \left( \sqrt{3} \ket{3} + \ket{9} \right).
\end{align}

This confirms the validity of the rotational construction and demonstrates a clear, experimentally feasible path to generating binomial codewords from a single resource state.

\subsection{Practical considerations}

The rotation-based construction of binomial codewords from a primitive state provides a unified and elegant theoretical framework, highlighting the symmetry underlying these codes. However, from an experimental perspective, this approach does not necessarily reduce the complexity of state preparation. The primitive states $\ket{\Theta}_{N,K}$ involve superpositions over all relevant Fock states $Nk$ for $k = 0, \dots, K$, whereas the binomial codewords occupy only a parity-restricted subset of this space.

Preparing such primitive states requires coherent control over a larger number of Fock components, increasing the sensitivity to decoherence and operational infidelity. Although in principle these states can be generated using the same MPJC interactions that are used to synthesize binomial codewords directly, doing so may be less efficient: if one already has access to controlled multiphoton interactions, generating the final codewords directly is typically more practical than preparing an intermediate superposition followed by rotation-based projections.

It is worth contrasting this with the family of cat codes, where the primitive state is a coherent state—a state that is straightforward to prepare experimentally with high fidelity. For cat codes, generating the entire family through rotations of this easily accessible primitive state is both conceptually elegant and experimentally practical~\cite{Li_AdvQTech_2023}. In contrast, the more complex primitive states required for binomial codes limit the practical advantage of the rotational construction in actual implementations.

Thus, while the rotational construction offers valuable conceptual insight, it does not intrinsically simplify the experimental path to realizing binomial code states.

\bibliography{references}

\end{document}